\documentclass{article}
\usepackage{amsmath,amsfonts,amssymb,setspace,comment,url}
\usepackage[a4paper]{geometry}
\usepackage{breqn,caption}
\usepackage{cite}
\usepackage{xcolor}
\usepackage{pifont}
\usepackage{hyperref}
\usepackage{bookmark}
\newcommand{\be}{\begin{equation}}
\newcommand{\ee}{\end{equation}}

\newcommand{\SU}[1]{\mathrm{SU}( #1 )}
\newcommand{\SL}[1]{\mathrm{SL}( #1 )}
\newcommand{\GL}[1]{\mathrm{GL}( #1 )}
\newcommand{\SO}[1]{\mathrm{SO}( #1 )}

\newcommand{\UO}{\mathrm{U}(1)}
\newcommand{\Spin}[1]{\mathrm{Spin}(#1)}

\newcommand{\USp}[1]{\mathrm{USp}(#1)}

\newcommand{\hK}{\hat{K}}
\newcommand{\K}{K}
\newcommand{\J}{J}
\newcommand{\hJ}{\hat{J}}

\newcommand{\mbf}[1]{\mathbf{#1}}
\newcommand{\gL}{\mathcal{L}}

\newcommand{\gM}{\mathcal{M}}

\numberwithin{equation}{section}

\newcommand{\Rsym}{(\SU{2}\times \UO)_R}
\newcommand{\SUt}{\SU{2}}
\newcommand{\coeff}{\frac{5^{1/6}\sqrt{2}}{R}}

\newcommand\Tstrut{\rule{0pt}{3ex}}         % = `top' strut
\newcommand\Bstrut{\rule[-1.3ex]{0pt}{0pt}}   % = `bottom' strut

\begin{document}

\begin{titlepage}
\vfill
\begin{flushright}
	HU-EP-20/45
\end{flushright}

\vfill

\begin{center}
	\baselineskip=16pt
	{\Large \bf Consistent truncations around half-maximal AdS$_5$\\ vacua of 11-dimensional supergravity}
	\vskip 2cm
	{\large \bf Emanuel Malek$^a$\footnote{\tt emanuel.malek@physik.hu-berlin.de},  Valent\'{i} Vall Camell$^{b}$\footnote{\tt valenti.vall-camell@ipht.fr}}
	\vskip .6cm
	{\it $^a$ Institut f\"ur Physik, Humboldt-Universit\"at zu Berlin, \\
	IRIS Geb\"aude, Zum Gro\ss en Windkanal 2, 12489 Berlin, Germany \\ \ \\
	\it $^b$ Institut de Physique Th\'eorique, Universit\'e Paris Saclay, CEA, CNRS,\\
	Orme des Merisiers, 91191
	Gif-sur-Yvette Cedex, France \\ \ \\}
	\vskip 1cm
\end{center}
\vfill

\begin{abstract}
We use exceptional field theory to systematically study all possible consistent truncations around half-maximal AdS$_5$ vacua of M-theory. We show that only truncations with at most three vector multiplets are consistent. The possible gaugings are SU(2)$\times$U(1) and ISO(3)$\times$U(1) where, in the first case, the U(1) factor can have different embeddings inside the global symmetry group SO$(5,n)$, where $n \leq 3$ equals the number of vector multiplets. We show that the ISO(3)$\times$U(1) truncation only exists around vacua that are locally an $S^4$ fibration over a Riemann surface, and argue that this is the only consistent truncation with vector multiplets around these vacua that can be constructed by going through the 7-dimensional SO(5) gauged supergravity obtained after an $S^4$ reduction. The consistent truncations with SU(2)$\times$U(1) gaugings with vector multiplets exist only if certain conditions are satisfied, which we derive.

\end{abstract}
\vskip 0cm

\vfill

\setcounter{footnote}{0} 

\end{titlepage}

\tableofcontents

\newpage

\section{Introduction}
A common problem in the study of string and supergravity compactifications is how to construct a lower-dimensional theory which captures relevant aspects of their dynamics. If there is a separation of scales, we can integrate out all massive modes above the compactification scale to obtain a lower-dimensional effective theory. However, without such a separation of scales, as for example in all well-understood AdS vacua\footnote{It is conjectured that no AdS vacua of string theory admit a separation of scales \cite{Lust:2019zwm}.} of string theory and supergravity, choosing a basis of modes to keep in the truncation is a difficult affair. As a minimum, we should require that the lower-dimensional theory obtained after truncation reproduces solutions to the original higher-dimensional supergravity. Such \textit{consistent truncations} used to be rare and difficult to construct \cite{Cvetic:2000dm}, with the only systematic construction arising for group manifolds, and isolated examples existing such as the $S^4$ \cite{deWit:1986oxb} and $S^7$ \cite{Nastase:1999cb,Nastase:1999kf} truncations of 11-dimensional supergravity. Despite these difficulties, consistent truncations have long played an important role in the AdS/CFT correspondence \cite{Aharony:2008ug} and as a solution-generating tool in supergravity.

Recently, the number of examples of consistent truncations and their systematic understanding has grown rapidly \cite{Hohm:2014qga,Lee:2014mla,Guarino:2015jca,Passias:2015gya,Lee:2015xga,Guarino:2015vca,Malek:2015hma,Baguet:2015iou,Ciceri:2016dmd,Cassani:2016ncu,Inverso:2016eet,Malek:2016bpu,Malek:2016vsh,Malek:2017cle,Malek:2017njj,Inverso:2017lrz,Hong:2018amk,Malek:2018zcz,Malek:2019ucd,Liu:2019cea,Cheung:2019pge,Cassani:2019vcl,Larios:2019lxq,Samtleben:2019zrh,Cassani:2020cod}. A key tool has been the development of Exceptional Field Theory (ExFT) \cite{Berman:2010is,Berman:2011cg,Berman:2012vc,Hohm:2013pua} and Exceptional Generalised Geometry (EGG) \cite{Pacheco:2008ps,Coimbra:2011ky,Coimbra:2012af}. ExFT and EGG are reformulations of supergravity theories where metric and flux degrees of freedom are treated on the same footing in a way that makes an $E_{d(d)}$ symmetry manifest\footnote{For the purposes of this paper, the differences between ExFT and EGG are not important and we will simply be referring to ExFT from now onwards.}. One of the benefits of this reformulation is that consistent truncations are now captured naturally and systematically. For example, consistent truncations preserving all supersymmetries are described by generalised Scherk-Schwarz truncations \cite{Aldazabal:2011nj,Geissbuhler:2011mx,Grana:2012rr,Berman:2012uy,Geissbuhler:2013uka,Berman:2013cli,Hohm:2014qga,Lee:2014mla} on ``generalised Leibniz parallelisable'' manifolds \cite{Lee:2014mla}. This has led to construction of the consistent truncation of IIB supergravity on $S^5$ \cite{Baguet:2015sma}, as well as a number of new consistent truncations on products of spheres and hyperboloids \cite{Hohm:2014qga,Lee:2014mla,Malek:2015hma,Baguet:2015iou,Ciceri:2016dmd,Cassani:2016ncu,Inverso:2016eet,Malek:2017cle,Samtleben:2019zrh}. On the other hand, consistent truncations preserving half the supersymmetries \cite{Malek:2016bpu,Malek:2017njj} in $11-d$ dimensions are described by a generalised $\Spin{d-1-n}$ structure, where $n$ labels the number of vector multiplets, and consistent truncations preserving a general number of supersymmetries were most recently described in \cite{Cassani:2019vcl}. Moreover, these tools led to a proof --  \cite{Malek:2017njj} for the half-maximal and subsequently \cite{Cassani:2019vcl} for the general supersymmetric case -- of the conjecture that every supersymmetric AdS vacuum of 10-/11-dimensional supergravity admits a consistent truncation keeping only the gravitational supermultiplets \cite{Gauntlett:2007ma}.

In the context of holography, it is especially useful to obtain consistent truncations around AdS vacua. These can then be used to study deformations of the AdS vacuum, such as those holographically modelling RG flows or constructing asymptotically AdS black holes. Therefore, it would be particularly useful to have a classification of possible consistent truncations around known AdS vacua. In \cite{Malek:2018zcz, Malek:2019ucd}, ExFT was used to systematically classify all possible consistent truncations around supersymmetric AdS$_6$ and AdS$_7$ vacua of 10-dimensional type IIB and IIA supergravity, respectively. In the present paper, we will extend this analysis to the case of AdS$_5$ vacua preserving 16 supercharges by classifying all their possible consistent truncations to half-maximal 5-dimensional gauged supergravity with arbitrary number of vector multiplets. Therefore, our work can be seen as a higher-dimensional completion of \cite{Louis:2015dca}.

This paper is organised as follows. In section \ref{sec:ExFT-review}, we review $E_{6(6)}$ exceptional field theory and how half-maximal AdS$_5$ vacua and consistent truncations around them are described within it. In section \ref{sec:AdS5-vacua}, we use this method to construct the generalised $G$-structures for all half-maximal AdS$_5$ vacua of 11-dimensional supergravity. This formulation is used in section \ref{sec:minimal-truncation} to construct consistent truncations keeping only the gravitational supermultiplet and in section \ref{sec:vector-multiplets} to systematically study all possible consistent truncations with vector multiplets around them. Finally, in section \ref{sec:AdS-SL(5)}, we analyse the particular case where the internal space is locally a $S^4$ fibration. In the remaining part of this section we give a summary of our results.

\subsection{Summary of results}
We summarise here the main results of the paper. In section \ref{sec:AdS5-vacua}, we construct the generalised $G$-structures for general AdS$_5$ geometries in 11-dimensional supergravity preserving 16 supercharges. These consist of the following warped product form
\begin{equation}
	\text{AdS}_5\times M_3\times S^2\times S^1,
\end{equation}
with $M_3$ a three dimensional manifold. These geometries are classified in terms of a function $D(x_1,x_2,y)$ satisfying the Toda equation \cite{Lin:2004nb}
\begin{equation}\label{eq:Toda-equation-(intro)}
	(\partial_{x_1}^2+\partial_{x_2}^2)\,D+\partial_y^2\,e^D=0 \,,
\end{equation}
with $(x_1, x_2, y)$ coordinates on $M_3$. In the process of describing these geometries by generalised $G$-structures, it is natural to define the following function and forms 
\begin{equation}
	\begin{split}
		p &= 2 \times 5^{1/6} \sqrt{2}\,R^2\, y \,,\qquad
		\nu_1 = 2 \times 5^{1/6} \sqrt{2}\,R^2\,  dx_1\,,\qquad 
		\nu_2 = 2 \times 5^{1/6} \sqrt{2}\,R^2\,  dx_2\,,\\
		d\chi &= -5^{1/6} \sqrt{2}\,R^2\,\Bigl(\epsilon^{ij}\partial_i D\, dx_j \wedge dy +\partial_y e^D\, dx_1\wedge dx_2\Bigr)\,,
	\end{split}
\end{equation}
where $R$ is the AdS$_5$ radius. Note that $d\chi$ can only be exact if $D$ satisfies the Toda equation \eqref{eq:Toda-equation-(intro)}.

With the generalised $G$-structures for the AdS$_5$ vacua, we can immediately construct their consistent truncations keeping only the gravitational supermultiplet. We do so in section \ref{sec:minimal-truncation} and show that this exactly reproduces the truncation of \cite{Gauntlett:2007sm}. In section \ref{sec:vector-multiplets}, we investigate the possibility to have consistent truncations keeping also vector multiplets. Our findings can be summarised as follows:
\begin{itemize}
	\item After analysing all possibilities, we conclude that truncations with at most 3 vector multiplets can be constructed around half-supersymmetric AdS$_5$ vacua. These organise into one of the following  (bosonic) representations of the R-symmetry group SU(2)$\times$U(1)
	\begin{equation}\label{eq:possible-reps-vec-mult}
		\textbf{1}_0 \,,\quad  \textbf{3}_0\,, \quad \textbf{1}_q \oplus \textbf{1}_{-q},
	\end{equation} 
	or into combinations thereof, as far as the total number of multiplets is not bigger than three.
	
	\item The truncation with one vector multiplet in the \textbf{1}$_0$ representation involve the following 1- and 2- forms
	\begin{equation}\label{eq:singlet-forms-intro}
		\begin{split}
		\bar\nu&=v_1\,\nu_1+v_2\,\nu_2+v_3\,dp\,,\\
		\bar\Phi&=\frac{1}{2\times 5^{1/6}\sqrt{2}\, R^2}\left[\left(v_1 \partial_y D-e^{-\frac{1}{2} D}\, v_3\, \partial_1 D\right)dp\wedge \nu_2  
		+ \left(-v_2 \partial_y D +e^{-\frac{1}{2} D}v_3 \partial_2 D\right) dp\wedge\nu_1\right]\\
		&\quad +v_2\,d\nu_1-v_1\, d\nu_2 + v_3 \, d\chi\,,
		\end{split}
	\end{equation}
	where the vector $v_1$, $v_2$ and $v_3$ are three functions of $M_3$ satisfying
	\begin{equation}\label{eq:norm-1-intro}
		v_1^2+v_2^2+v_3^2=1\,.
	\end{equation}
	The truncation exists if one can find three functions $(v_1,v_2,v_3)$ satisfying the above condition such that the forms $\bar\nu$ and $\bar\Phi$ are closed,
	\begin{equation}
		d\bar\nu=d\bar\Phi=0\,.
	\end{equation}
	
	\item The consistent truncation with 3 vector multiplets transforming in the \textbf{3}$_0$ exists if and only if the AdS$_5$ vacuum comes from a $S^4$ warped over a Riemann surface. This truncation was first constructed in \cite{Cheung:2019pge} and arises by truncating 11-dimensional supergravity first on $S^4$ to 7-dimensional supergravity.

	\item The truncation with two vector multiplets in the \textbf{1}$_q$ $\oplus$ \textbf{1}$_{-q}$ representation involves two copies of the forms \eqref{eq:singlet-forms-intro}, which we call $(\bar\nu_1,\bar\Phi_1)$ and $(\bar\nu_2,\bar\Phi_2)$, and are characterised by the functions $(v_1,v_2,v_3)$ and $(w_1,w_2,w_3)$ respectively. Apart from satisfying \eqref{eq:norm-1-intro} independently, they also need to satisfy the condition
	\begin{equation}\label{eq:perpendicular-cond-vecmult-intro}
		v_1\,w_1+v_2\,w_2+v_3\,w_3=0\,.
	\end{equation} 
	In this case, the truncation exists if we can find functions for which the conditions
	\begin{equation}
		\bar\Phi_1=\frac{1}{q}\,d\bar\nu_2\,,\qquad \bar\Phi_2=-\frac{1}{q}\,d\bar\nu_1\,,
	\end{equation}
	where $q\in\mathbb{Z}$ is the U(1) charge, are satisfied.
	
	\item Finally, to have truncations with different combinations of the representations \eqref{eq:possible-reps-vec-mult}, we need
	\begin{enumerate}
		\item That truncations with vector multiplets in the individual representations exist, and
		\item That all forms of the type \eqref{eq:singlet-forms-intro} involved in the individual representations satisfy conditions of the form \eqref{eq:perpendicular-cond-vecmult-intro} pairwise. 
	\end{enumerate}
\end{itemize}
In table \ref{table:vector-mult-summary} we summarise all these possibilities.
	
In section \ref{sec:AdS-SL(5)}, we prove that the ISO$(3) \times \UO$ truncation with 3 vector multiplets and the minimal truncation are the only possible consistent truncations around half-maximal AdS$_5$ vacua of $\SO{5}$ 7-dimensional gauged supergravity.

\begin{table}
	\begin{center}
		\begin{tabular}{|c|c|c|c|}
			\hline
			$n$ & $\Rsym$ rep. & Conditions to be satisfied&Gauging \Tstrut\Bstrut\\
			\hline
			1 & 
			\textbf{1}$_0$ & $d\bar\nu=d\bar\Phi=0$&SU(2)$\times$U(1)
			\Tstrut\Bstrut\\
			\hline
			2 & 
			\textbf{1}$_0$ $\oplus$ \textbf{1}$_0$ & Two different \textbf{1}$_0$'s  
			satisfying \eqref{eq:perpendicular-cond-vecmult-intro}& SU(2)$\times$U(1)
			\Tstrut\Bstrut \\
			\hline
			2 & 
			\textbf{1}$_q$ $\oplus$ \textbf{1}$_{-q}$ & $\bar\Phi_1=\frac{1}{q}\,d\bar\nu_2\quad \bar\Phi_2=-\frac{1}{q}\,d\bar\nu_1$&SU(2)$\times$U(1)$'$\Tstrut\Bstrut \\
			\hline
			3 & \textbf{1}$_0$ $\oplus$ \textbf{1}$_0$ $\oplus$ \textbf{1}$_0$
			& Three different \textbf{1}$_0$'s 
			satisfying \eqref{eq:perpendicular-cond-vecmult-intro} pairwise&SU(2)$\times$U(1)
			\Tstrut\Bstrut\\
			\hline 
			3 & \textbf{1}$_0$ $\oplus$ \textbf{1}$_q$ $\oplus$ \textbf{1}$_{-q}$ 
			& \textbf{1}$_0$ and \textbf{1}$_q$ $\oplus$ \textbf{1}$_{-q}$ 
			satisfying \eqref{eq:perpendicular-cond-vecmult-intro} pairwise&SU(2)$\times$U(1)$'$
			\Tstrut\Bstrut\\
			\hline 
			3 & 
			\textbf{3}$_0$ & $S^4$ fibred over Riemann surface & ISO(3)$\times$U(1) \Tstrut\Bstrut\\
			\hline 
		\end{tabular}
	\end{center}
	\caption{\label{table:vector-mult-summary} Summary of the possible consistent truncations with $n$ vector multiplets around half-maximal AdS$_5$ vacua and their gaugings. The prime on $\UO$ denotes a non-standard embedding of the $\UO \subset \SO{5,n}$. The ``standard embedding'' corresponds to $\UO \subset \SO{5} \subset \SO{5,n}$, whereas $\UO'$ indicates a linear combination of $\UO \subset \SO{5}$ and $\UO \subset \SO{n}$ which can be read off from \eqref{eq:gauging}. The $n=1$ case is characterised by the forms $\bar\nu$ and $\bar\Phi$ given in \eqref{eq:singlet-forms-intro} satisfying \eqref{eq:norm-1-intro}. The  $n=2$ case transforming in the \textbf{1}$_q$ $\oplus$ \textbf{1}$_{-q}$  representation is characterised by two pairs of forms, $(\bar\nu_1, \bar\Phi_1)$ and $(\bar\nu_2, \bar\Phi_2)$, both of the form \eqref{eq:singlet-forms-intro} and satisfying individually \eqref{eq:norm-1-intro} together with \eqref{eq:perpendicular-cond-vecmult-intro}. Consistent truncations with $n>3$ do not exist.}
\end{table}

\section{$E_{6(6)}$ ExFT and consistent truncations around AdS vacua}\label{sec:ExFT-review}
In this section, we review the relevant aspects of $E_{6(6)}$ exceptional field theory \cite{Hohm:2013vpa} needed for the constructions discussed in the upcoming sections. This theory is a reformulation of 11-dimensional (10-dimensional) supergravity based on a split into 5 external and 6 (5 in the case of 10-dimensional supergravity) internal directions. The internal degrees of freedom are organised into representations of the exceptional group $E_{6(6)}$. The internal coordinates themselves are embedded into a larger set of generalised coordinates transforming in the $\textbf{27}$ representation of $E_{6(6)}$, but subject to the ``section condition''
\begin{equation} \label{eq:SectionCondition}
	d^{MNP}\partial_N\otimes\partial_P = 0\,,
\end{equation}
where $M = 1, \ldots, 27$, $d_{MNP}$ is the totally symmetric cubic invariant of $E_{6(6)}$ and the derivatives act on any pair of fields or gauge parameters in the theory. This constraint is solved by restricting the dependence of all fields to a subset of the generalised coordinates, breaking the $E_{6(6)}$ symmetry to some smaller subgroup. Upon solving the section condition in this way, ExFT reduces to 11-dimensional or type II supergravity.

The internal diffeomorphisms and gauge transformations are unified into the notion of generalised diffeomorphism along the generalised internal coordinates. Its action, parametrised by a gauge parameter $\Lambda\in\textbf{27}$ acting on any generalised vector $V\in\textbf{27}$, is given by the generalised Lie derivative\footnote{Here and throughout this paper we will take our generalised vector fields to have weight $1/3$.}
\begin{equation} \label{eq:GenLieDerivative}
	\mathcal{L}_{\Lambda}V^M=\Lambda^N\partial_N V^M-V^N\partial_N \Lambda^M+10\, d_{NLP}d^{PMQ}\,V^L\partial_Q\Lambda^N \,.
\end{equation}
Using this definition, the algebra of generalised diffeomorphism closes only upon solving the section condition \eqref{eq:SectionCondition}.

The field content of $E_{6(6)}$ ExFT is given by\cite{Hohm:2013vpa}
\begin{equation}\label{eq:ExFT-field-content}
	\{g_{\mu\nu}\,,\quad \mathcal{A}_\mu{}^N\,, \quad \mathcal{B}_{\mu\nu,\, N}\,,\quad \mathcal{M}_{MN}\}\,, \qquad \mu = 0, \ldots, 4 \,,
\end{equation}
where the external metric $g_{\mu\nu}$ is a $E_{6(6)}$ singlet, the one-form $\mathcal{A}_\mu{}^N$ transforms in the \textbf{27} representation and the two-form $\mathcal{B}_{\mu\nu,\, N}$ in the $\overline{\textbf{27}}$ representation of $E_{6(6)}$. All purely internal degrees of freedom are encoded into the generalised metric $\mathcal{M}_{MN}$, that parametrise the coset space $E_{6(6)}/$USp(8). 

Finally, throughout this paper, we will make use of the following $E_{6(6)}$-invariant algebraic operations: given two vectors in the \textbf{27} representation, we define a wedge product $\wedge$ as the algebraic map \cite{Malek:2017njj} (the same operation is denoted by $\bullet$ in \cite{Wang:2015hca})
\begin{equation}\label{eq:wedge27-def}
	\begin{split}
		\textbf{27}\otimes\textbf{27}&\longrightarrow \overline{\textbf{27}} \,, \\
		(V_1,\, V_2) &\longrightarrow (V_1\wedge V_2)_M =d_{MNP}V_1{}^NV_2{}^P \,,
	\end{split}
\end{equation}
where $d_{MNP}$ is the $E_{6(6)}$ invariant. Furthermore, we define the $\wedge$ product between a \textbf{27} and a $\overline{\textbf{27}}$ representations as
\begin{equation}\label{eq:wedge27bar-def}
	\begin{split}
		\textbf{27}\otimes\overline{\textbf{27}}&\longrightarrow \textbf{1} \,, \\
		(V, W) &\longrightarrow V\wedge W=V^NW_N \,.
	\end{split}
\end{equation}
By choosing a solution to the section constraint, these wedge products decompose into standard wedge products between differential forms, as we review in appendix \ref{sec:app-11d-section}.

\subsection{Half-supersymmetric AdS$_5$ vacua in $E_{6(6)}$ ExFT} \label{s:AdS5Str}
Supersymmetric vacua are naturally described in ExFT in terms of generalised G-structures. In order for $M_5\times M_{int}$ to be a half-supersymmetric vacuum, the internal space $M_{int}$ has to admit 16 no-where vanishing real Killing spinor fields. As shown in \cite{Malek:2017njj,Malek:2016vsh,Malek:2016bpu}, this is equivalent to an algebraic and differential condition on $M_{int}$.

The algebraic condition is that $M_{int}$ must admit a generalised USp(4) structure, defined by six nowhere vanishing generalised vectors $\{J_u, \hat K\}$, with $u=1\dots 5$, transforming in the \textbf{27} representation of $E_{6(6)}$ and satisfying the algebraic constraints 
\begin{equation}\label{eq:SO5-structure-cond}
	\begin{split}
		J_u\wedge J_v &=\frac{1}{5}\, \delta_{uv}\, J_w \wedge J^w \,, \\
		\hK\wedge\hK & = 0 \,, \\
		\hK\wedge J_w \wedge J^w  & > 0 \,,
	\end{split}
\end{equation}
where $u$ indices are raised/lowered by $\delta_{uv}$ and the wedge products between generalised objects are defined in \eqref{eq:wedge27-def}-\eqref{eq:wedge27bar-def}. The maximal commutant of USp(4) $\subset E_{6(6)}$ is the R-symmetry group USp(4)$_R$, that rotates the structures $J_u$ among themselves and leaves $\hat{K}$ invariant. For the upcoming discussion, it is useful to define the following fields
\begin{equation}
	K=\frac{1}{5}\,J_u\wedge J^u\,,\qquad \kappa^3=K\wedge\hat K \,,
\end{equation}
where $K$ is an object in the $\overline{\textbf{27}}$ representation of $E_{6(6)}$ and $\kappa$ a density of weight $1/3$. Note that the generalised vector fields $J_u$ and $\hK$ in \eqref{eq:SO5-structure-cond} have weight $1/3$ and thus $K$ has weight $2/3$.

In order to describe an AdS vacuum, the structure defined by \eqref{eq:SO5-structure-cond} needs to be \textit{weakly integrable} \cite{Malek:2017njj}, which is equivalent to demanding that the internal geometry satisfies the corresponding BPS conditions. This condition implies the following differential constraints
\begin{equation}
	\begin{split}
		\gL_{J_u}J_v&=\Lambda_{uvw}J^w\,,\\
		\gL_\hK J_u&=\frac{1}{3!}\,\epsilon_{uvwxy}\Lambda^{vwx}J^y\,, \\
		\gL_{J_u} \hK &= 0 \,,
	\end{split}
\end{equation}
where $\Lambda_{uvw}=\Lambda_{[uvw]}$ is a constant totally antisymmetric tensor that  encodes the information of the cosmological constant. Using SO(5) rotations, one can always bring the constant tensor $\Lambda$ into a form where only the component $\Lambda_{123}$ and its permutations are non-zero. We observe that the presence of such tensor breaks the  R-symmetry group USp(4)$_R$ into (SU(2)$\times$U(1))$_R$, which is the R-symmetry group preserving the AdS$_5$ vacuum, or equivalently the R-symmetry group of the holographically dual superconformal field theory. By splitting the index $u$ into $(A,\,i)$ with $A=1,2,3$ being the SU(2) index and $i=1,2$ the U(1) one, we can now write
\begin{equation}
	\Lambda_{ABC} = - \frac{5^{1/6} \sqrt{2}}{R} \epsilon_{ABC} \,.
\end{equation}
Here we have fixed the conventions such that $R$ is the radius of the AdS$_5$ vacuum. The weakly integrable conditions now become
\begin{equation}\label{eq:Gen-weakly-integrability-cond}
	\begin{split}
		\gL_{J_A}J_B&=-\frac{5^{1/6}\sqrt{2}}{R}\,\epsilon_{ABC}\,J^C\,,\\
		\gL_\hK J_i&=-\frac{5^{1/6}\sqrt{2}}{R}\, \epsilon_{ij}\,J^j\,,\\
		\gL_\hK J_A&=\gL_{J_A}J_i=\gL_{J_i}J_A=\gL_{J_i}J_j=\gL_{J_u} \hK=0\,.
	\end{split}
\end{equation}

\subsection{Generalised metric from half-maximal structures}
So far we have described the half-maximal AdS$_5$ vacua in terms of a generalised $\USp{4}$ structure \eqref{eq:SO5-structure-cond}. However, the generalised $\USp{4}$ structure also defines a generalised metric $\gM$, from which the supergravity background fields can be read off using the dictionaries described in appendix \ref{sec:app-11d-section}. For example, the generalised metric was constructed out of the half-maximal structures in $\SO{5,5}$ and $\SL{5}$ ExFT in \cite{Malek:2018zcz}. The analogous expression in $E_{6(6)}$ can be found by constructing a $\USp{4}_R$-invariant combination of $\J_u$ and $\hK$ which lives in the coset $E_{6(6)}/\USp{8}$. Here we will take
\begin{equation}\label{eq:Gen-Metric-from-structures}
	\begin{split}
		\gM_{MN}=2\times 5^{1/3}&\left(
		\frac{1}{\kappa^4}\,\hJ_{u,M}\hJ^u{}_N
		+\frac{1}{\kappa^4}K_M K_N
		-\frac{1}{2\kappa}d_{MNP}\hK^P\right.\\
		&\qquad\left.-\frac{10}{3\kappa^7}\epsilon^{u_1\dots u_5}(d\cdot J_{u_1})_{M P}(d\cdot \hJ_{u_2})^{PQ}(d\cdot J_{u_3})_{QR}(d\cdot \hJ_{u_4})^{RS}(d\cdot J_{u_5})_{SN}\right) \,,
	\end{split}
\end{equation}
where we defined
\begin{equation}
	\begin{split}
		\hJ_{u,M} &= d_{MNP}J_u{}^N\hK^P \,, \\
		(d\cdot J_u)_{MN}&=d_{MNP}J_u{}^P \,, \\
		(d\cdot \hJ_u)^{MN}&=d^{MNP}\hJ_{u,P} \,.
	\end{split}
\end{equation}
The warp factor of the 5-dimensional metric is then given by
\begin{equation} \label{eq:Warp}
	f_1= |\det g|^{-1/3}\kappa^2\,,
\end{equation}
where $g$ is the internal metric that can be read off from \eqref{eq:Gen-Metric-from-structures} and the dictionary to 11-dimensional supergravity given in appendix \ref{sec:app-11d-section}.

Note that the coefficients in \eqref{eq:Gen-Metric-from-structures} and \eqref{eq:Warp} are not unique since $J_u$ and $\hK$ can be rescaled, whilst still satisfying \eqref{eq:SO5-structure-cond}. Indeed, in \cite{Cassani:2019vcl}, the generalised metric was given with different coefficients to our \eqref{eq:Gen-Metric-from-structures}. As a result of such a rescaling, the coefficients in the differential conditions \eqref{eq:Gen-weakly-integrability-cond} would change. However, the coefficients in \eqref{eq:Gen-Metric-from-structures} become unique once we impose the differential conditions \eqref{eq:Gen-weakly-integrability-cond}.

\subsection{Consistent truncations around half-maximal AdS$_5$ vacua}
As shown in \cite{Malek:2017njj}, a consistent truncation of 11-dimensional/10-dimensional maximal supergravity to a half-maximally supersymmetric supergravity in five dimensions is described in a similar way to the AdS$_5$ vacua reviewed in section \ref{s:AdS5Str}. Here we focus on consistent truncations around half-maximal AdS$_5$ vacua, whose construction we review in the following.

\subsubsection{Minimal consistent truncation} \label{sec:minimal-truncation-review}
Given any half-maximal AdS$_5$ vacuum, we can immediately construct a minimal consistent truncation to half-maximal gauged supergravity with $\SU{2} \times \UO$ gauging \cite{Romans:1985ps} and containing only the gravitational supermultiplet using the following truncation Ansatz
\begin{equation}
	\begin{split} \label{eq:MinimalTruncation}
		\mathcal{J}_u{}^M(x,Y)&=X(x)J_u{}^M(Y)\,,\\
		\hat{\mathcal{K}}^M(x,Y)&=X^{-2}(x)\hK^M (Y)\,,
	\end{split}	
\end{equation}
for the scalar sector and
\begin{equation}
	\begin{split} \label{eq:MinimalTruncationTensors}
		{\cal A}_\mu{}^M(x,Y) &= A_\mu{}^u(x) J_u{}^M(Y) + A_\mu{}^0(x) \hK^M(Y) \,, \\
		{\cal B}_{\mu\nu\,M}(x,Y) &= - B_{\mu\nu}{}^u(x) \hJ_{u,M}(Y) - B_{\mu\nu}(x) K_M(Y) \,, \\
		g_{\mu\nu}(x,Y) &= \bar{g}_{\mu\nu}(x) \kappa^{2/3}(Y) \,,
	\end{split}
\end{equation}
for the other fields fo the $E_{6(6)}$ ExFT. Here, $Y^M$ denote the internal coordinates on $M_{int}$ and $x^\mu$ the external coordinates on $M_5$, $X$ is the scalar field, $\left(A_\mu{}^u, A_\mu{}^0\right)$ are the vector fields, $\left(B_{\mu\nu}{}^u, B_{\mu\nu}{}^0\right)$ the tensor fields and $\bar{g}_{\mu\nu}$ the metric of the 5-dimensional gravitational supermultiplet. The ExFT generalised metric is then computed using ${\cal J}_u$ and ${\cal \hK}$ instead of $J_u$ and $\hK$ in \eqref{eq:Gen-Metric-from-structures}.

\subsubsection{Consistent truncation with vector multiplets} \label{sec:vector-multiplets-review}
Moreover, for special half-maximal AdS$_5$ vacua, we can construct a consistent truncation with vector multiplets. This arises when the AdS$_5$ vacuum admits a generalised Spin$(5-n) \subset$ USp$(4)$ structure, where $n$ labels the number of vector multiplets obtained in the 5-dimensional gauged supergravity, and we define Spin$(1) =$ Spin$(0) = \mathbb{Z}_2$. A generalised Spin$(5-n)$ structure is equivalent to $M_{int}$ admitting a further $n$ generalised vector fields, $\bar{J}_{\bar{u}}$, $\bar{u}=1,\dots, n$, satisfying the algebraic conditions
\begin{equation}\label{eq:vec-mult-algebraic-cond-rev}
	\begin{split}
		\bar{J}_{\bar{u}}\wedge\bar{J}_{\bar{v}}&=-\delta_{\bar{u}\bar{v}} K \,, \\
		\bar{J}_{\bar{u}}\wedge J_u&=0\,.
	\end{split}
\end{equation}
Note that this immediately implies that we can only construct truncations with at most $n\le 5$ vector multiplets.

Moreover, for the truncation around the AdS vacuum to be consistent, the following differential conditions must be satisfied
\begin{equation}\label{eq:vec-mult-diff-cond-rev}
	\begin{split}
		\gL_{J_u}\bar{J}_{\bar{v}}&=-f_{u\bar{v}}{}^{\bar{w}}\bar{J}_{\bar{w}}\,,\\
		\gL_{\bar{J}_{\bar{u}}}\bar{J}_{\bar{v}}&=-f_{\bar{u}\bar{v}}{}^w J_w-f_{\bar{u}\bar{v}}{}^{\bar{w}}\bar{J}_{\bar{w}}\,,\\
		\gL_{\bar{J}_{\bar{u}}}\hK&=0\,,\\
		\gL_\hK\bar{J}_u&=-\xi_{\bar{u}}{}^{\bar{v}}\bar{J}_{\bar{v}}\,,
	\end{split}
\end{equation}
where $\xi_{\bar{u}}{}^{\bar{v}}$, $f_{u\bar{v}}{}^{\bar{w}}$, $f_{\bar{u}\bar{v}}{}^w$ and $f_{\bar{u}\bar{v}}{}^{\bar{w}}$ are constants. In particular, the first and last equations of \eqref{eq:vec-mult-diff-cond-rev} imply that the extra generalised vector fields $J_{\bar{u}}$ corresponding to the vector multiplets must form a representation of the $\left(\SU{2} \times \UO\right)_R$ symmetry group of the AdS$_5$ vacuum. This results in a gauged supergravity with embedding tensor \cite{Schon:2006kz} whose components can be read off as explained in \cite{Malek:2017njj}. Thus, we find
\begin{equation}
	f_{uvw} = - \Lambda_{uvw} \,, \qquad \tilde{f}_{uv} = - \frac{1}{3!} \epsilon_{uvwxy} \Lambda^{wxy} \,,
\end{equation}
are defined by the vacuum \eqref{eq:Gen-weakly-integrability-cond} and the remaining non-vanishing components are defined by \eqref{eq:vec-mult-diff-cond-rev}. The embedding tensor defines the gauging of the supergravity via the gauge-covariant derivative
\begin{equation} \label{eq:gauging}
	D_\mu = \nabla_\mu - A_\mu{}^a f_{a}{}^{bc}\, t_{bc} - A_\mu{}^0\, \xi^{bc}\, t_{bc} \,,
\end{equation}
where $a, b = 1, \ldots, 5+n$ are raised/lowered by the $\SO{5,n}$ metric
\begin{equation}
	\eta_{ab} = \begin{pmatrix}
		\delta_{uv} & 0 \\
		0 & - \delta_{\bar{u}\bar{v}}
	\end{pmatrix} \,,
\end{equation}
and $t_{ab} = t_{[ab]}$ are the $\SO{5,n}$ generators.

For the consistent truncation Ansatz, it is useful to define the $5+n$ generalised vector fields
\begin{equation}
	J_a{}^M = \left(J_u{}^M,\, J_{\bar{u}}{}^M\right) \,,
\end{equation}
as well as
\begin{equation}
	\hJ_a = J_a \wedge \hK \,.
\end{equation}
Now the consistent truncation Ansatz is given by
\begin{equation} \label{eq:VecTruncAnsatz}
	\begin{split}
		\mathcal{J}_u{}^M(x,Y)&=X(x)\, b_u{}^{a}(x) J_a{}^M(Y)\,,\\
		\hat{\mathcal{K}}^M(x,Y)&=X^{-2}(x) \hK^M(Y)\,,\\
		{\cal A}_\mu{}^M(x,Y) &= A_\mu{}^a(x) J_a{}^M(Y) + A_\mu{}^0(x) \hK^M(Y) \,, \\
		{\cal B}_{\mu\nu\,M}(x,Y) &= - B_{\mu\nu}{}^a(x) \hJ_{a,M}(Y) - B_{\mu\nu}(x) K_M(Y) \,, \\
		g_{\mu\nu}(x,Y) &= \bar{g}_{\mu\nu}(x) \kappa^{2/3}(Y) \,,
	\end{split}
\end{equation}
where the $x$-dependent fields are the fields of the 5-dimensional gauged supergravity. As explained in \cite{Malek:2017njj}, the $b_u{}^a$ are constrained by
\begin{equation}
	b_u{}^a b_v{}^b \eta_{ab} = \delta_{uv} \,,
\end{equation}
and are identified by SO$(5)$ rotations on the $u, v$ indices. Thus, the $b_u{}^a \in \frac{\SO{5,n}}{\SO{5}\times\SO{n}}$, together with $X \in \mathbb{R}^+$ parameterise the scalar coset space of the 5-dimensional half-maximal gauged supergravity.

\section{Half maximal AdS$_5$ vacua from 11d supergravity}\label{sec:AdS5-vacua}
The general half-maximal AdS$_5$ vacua of M-theory take the local form AdS$_5 \times S^2 \times S^1 \times M_3$, where $M_3$ is a 3-dimensional compact manifold with boundaries. In order to study their consistent truncations, we must first describe these backgrounds in terms of generalised $\USp{4}$ structures. Using the decomposition of ExFT fields into 11-dimensional supergravity discussed in appendix \ref{sec:app-11d-section}, the most general Ansatz for the structures on $S^2 \times S^1 \times M_3$ compatible with the $\Rsym$ symmetry is
\begin{equation} \label{eq:VecTruncationAnsatz}
	\begin{split}
		J_A&= \frac{5^{1/6}\sqrt{2}}{R}\, v_A+\\
		&\quad+ \Bigl(a\,Y_A\,vol_{S^2}+p\, dY_A\wedge vol_{S^1}+f \,\theta_A\wedge vol_{S^1} +\chi\wedge dY_A+\xi\wedge \theta_A+Y_A\, \zeta\wedge vol_{S^1}+Y_A\,\Xi\Bigr)\\
		&\quad+\Bigl(Y_A\,\Psi \wedge vol_{S^2}\wedge vol_{S^1}+l\,Y_A\,vol_3 \wedge vol_{S^2} +m\,vol_3\wedge dY_A\wedge vol_{S^1} +n\,vol_3\wedge \theta_A\wedge vol_{S^1}\Bigr)\,,\\
		J_i&=\Bigl(b_1\,w_i\,vol_{S^2}+b_2\,\epsilon w_i\,vol_{S^2}+w_i\,\Phi_1+\epsilon w_i\,\Phi_2+w_i\,\nu_2\wedge vol_{S^1}+\epsilon w_i\,\nu_1\wedge vol_{S^1}\Bigr)\\
		&\quad+\Bigl( w_i\,\Pi_1\wedge vol_{S^2}\wedge vol_{S^1}+\epsilon w_i\,\Pi_2\wedge vol_{S^2}\wedge vol_{S^1}+ z_1\,w_i\, vol_3\wedge vol_{S^2}+ z_2\,\epsilon w_i\, vol_3\wedge vol_{S^2}\Bigr) \,, \\
		\hat{K}&=\frac{5^{1/6}\sqrt{2}}{R}\, v_{S^1}+(c\, vol_{S^2}+\Omega+\psi\wedge vol_{S^1})+(\Theta\wedge vol_{S^2}\wedge vol_{S^1}+e\,vol_3\wedge vol_{S^2}) \,,
	\end{split}
\end{equation}
where small Latin, small Greek and capital Greek letters represent functions, 1-forms and 2-forms on $M_3$ respectively. The forms $vol_3$, $vol_{S^2}$ and $vol_{S^1}$ are the volume forms on $M_3$, $S^2$ and $S^1$. $Y_A$, $w_i$ are the embedding coordinates of $S^2 \hookrightarrow \mathbb{R}^3$ and $S^1 \hookrightarrow \mathbb{R}^2$, respectively, satisfying
\begin{equation}
	Y_A Y_B \delta^{AB} = w_i w_j \delta^{ij} = 1 \,,
\end{equation}
and where we raise/lower the $A, B = 1, \ldots, 3$ and $i, j = 1, 2$ indices with $\delta_{AB}$ and $\delta_{ij}$, respectively. We also defined the Killing vectors on the round $S^2$ and $S^1$ as $v_A$ and $v_{S^1}$, respectively, as well as $\theta_A \equiv \star dY_A$, the Hodge dual with respect to the round $S^2$, and $\epsilon w_i \equiv \epsilon_{ij} w^j$. The details of all these objects can be found in appendix \ref{sec:app-spheres}.  

The algebraic conditions \eqref{eq:SO5-structure-cond} impose
\begin{equation}\label{eq:AdS5structures-1}
	\begin{split}
		J_A&= \frac{5^{1/6}\sqrt{2}}{R}\, v_A+ \Bigl(p\, dY_A\wedge vol_{S^1}+\chi\wedge dY_A+Y_A\, \zeta\wedge vol_{S^1}+Y_A\,\Xi\Bigr)\\
		&\quad+\frac{R}{5^{1/6}\sqrt{2}}\, \Bigl(Y_A\,(p\,\Xi-\chi\wedge\zeta) \wedge vol_{S^2}\wedge vol_{S^1}-\,Y_A\,\chi\wedge \Xi \wedge vol_{S^2} +m\,vol_3\wedge dY_A\wedge vol_{S^1}\\
		&\qquad\qquad\qquad +\zeta\wedge\Xi\wedge \theta_A\wedge vol_{S^1}\Bigr)\,,\\
		J_i&=\Bigl(w_i\,\Phi_1+\epsilon w_i\,\Phi_2+w_i\,\nu_2\wedge vol_{S^1}+\epsilon w_i\,\nu_1\wedge vol_{S^1}\Bigr)\\
		&\quad+\frac{R}{5^{1/6}\sqrt{2}}\, \Bigl( w_i\,(p\,\Phi_1-\chi\wedge\nu_2)\wedge vol_{S^2}\wedge vol_{S^1}+\epsilon w_i\,(p\,\Phi_2-\chi\wedge\nu_1)\wedge vol_{S^2}\wedge vol_{S^1}\\
		&\qquad\qquad\qquad -w_i\,\Phi_1\wedge\chi\wedge vol_{S^2} -\,\epsilon w_i\, \Phi_2\wedge\chi\wedge vol_{S^2}\Bigr) \,, \\
		\hat{K}&=\frac{5^{1/6}\sqrt{2}}{R}\, v_{S^1}+(c\, vol_{S^2}+\Omega)+\left( \frac{R}{5^{1/6}\sqrt{2}}\,c\,\Omega\wedge vol_{S^2}\wedge vol_{S^1}+e\,vol_3\wedge vol_{S^2}\right) \,,
	\end{split}
\end{equation}
where the 1-forms $\zeta$, $\nu_1$, $\nu_2$ and the 2-forms $\Xi$, $\Phi_1$, $\Phi_2$ satisfy the constraints 
\begin{equation}\label{eq:alg-cond-vacua}
	\begin{split}
		\zeta\wedge\Phi_2=-\nu_1\wedge\Xi\,, \qquad
		\zeta\wedge\Phi_1&=-\nu_2\wedge\Xi\,,\qquad
		\nu_1\wedge\Phi_1=-\nu_2\wedge\Phi_2 \,, \\ \nu_2\wedge\Phi_1=\nu_1&\wedge\Phi_2=\zeta\wedge\Xi\,,
	\end{split}
\end{equation}
and $p$, $m$, $c$  and $e$ are at this stage unconstrained functions. The field $K=\frac{1}{5}J_u\wedge J^u$ is 
\begin{equation}
	\begin{split}
		K &= \sqrt{\frac{2}{5}}\Bigl(\, \zeta\wedge\Xi\wedge vol_{S^1} +\frac{R}{5^{1/6}\sqrt{2}}\, (\chi-p\, vol_{S^1})\otimes\zeta\wedge\Xi\wedge vol_{S^2}\wedge vol_{S^1}\Bigr)\,,
	\end{split}
\end{equation}
and
\begin{equation}
	K\wedge\hK=\sqrt{\frac{2}{5}}\,(c-p)\,\zeta\wedge\Xi\wedge vol_{S^1}\wedge vol_{S^2}>0 \,,
\end{equation}
where the last inequality follows from \eqref{eq:SO5-structure-cond}. Allowing the $S^2$ and $S^1$ to shrink at the boundaries of $M_3$, we have
\begin{equation}\label{eq:pos-condition-vacua-1}
	(c-p)\,\zeta\wedge\Xi \ge 0\,,
\end{equation}
with equality on the boundaries.

For later purposes, it is worth noticing that the constraints \eqref{eq:alg-cond-vacua} together with the condition \eqref{eq:pos-condition-vacua-1} imply that the 1-forms $\nu_1$, $\nu_2$ and $\zeta$ are non-vanishing and linear independent at each point of $M_3$ except for the boundaries. To show this, let us start by considering the situation where $\zeta= a\, \nu_1$ for some no-where-vanishing function $a$. Then, from \eqref{eq:alg-cond-vacua},
\begin{equation}
	\frac{1}{a}\,\zeta\wedge\Xi=\nu_1\wedge\Xi=-\zeta\wedge\Phi_2=-a\,\nu_1\wedge\Phi_2 =-a\, \zeta\wedge\Xi 
	\quad\Rightarrow\quad 
	\left(\frac{1}{a}+a\right)\zeta\wedge\Xi=0\,,
\end{equation}
contradicting \eqref{eq:pos-condition-vacua-1}. Using analogous arguments, it is straightforward to show that none of the three 1-forms can be proportional pairwise. Next, suppose that $\zeta= a\, \nu_1+b\,\nu_2$ with $a$ and $b$ no-where-vanishing functions. Plugging this into the left- and right-hand sides of the first three conditions in \eqref{eq:alg-cond-vacua} and using the last two, one finds the following conditions
\begin{equation}
	\begin{split}
		\left(a+\frac{1}{a}\right)\,\zeta\wedge\Xi +b\,\left(\nu_2\wedge\Phi_2-\frac{1}{a}\,\nu_2\wedge\Xi\right)&=0\,,\\
		\left(b+\frac{1}{b}\right)\,\zeta\wedge\Xi +a\,\left(\nu_1\wedge\Phi_1-\frac{1}{b}\,\nu_1\wedge\Xi\right)&=0\,,\\
		\left(\frac{a}{b}+\frac{b}{a}\right)\,\zeta\wedge\Xi -\,\left(\frac{1}{a}\,\zeta\wedge\Phi_1+\frac{1}{b}\,\zeta\wedge\Phi_2\right)&=0\,.
	\end{split}
\end{equation}
Adding them with appropriate factors and using again \eqref{eq:alg-cond-vacua} one obtains
\begin{equation}
	(a^2+b^2+1)\,\zeta\wedge\Xi=0\,,
\end{equation}
contradicting again \eqref{eq:pos-condition-vacua-1}, which concludes the proof. The fact that the 1-forms are linear independent at all points of the manifold implies that
\begin{equation}\label{eq:linear-idep-cond-1}
	\zeta\wedge\nu_1\wedge\nu_2\neq 0\,,
\end{equation}
everywhere except the boundaries. 

The differential conditions \eqref{eq:Gen-weakly-integrability-cond} imposed on \eqref{eq:AdS5structures-1} imply that
\begin{equation}\label{eq:diff-cond-vacua}
	\Xi=-d\chi\,,\qquad \zeta=dp\,,\qquad \Phi_1=d\nu_1\,,\qquad\Phi_2=-d\nu_2\,,
\end{equation}
as well as
\begin{equation} \label{eq:dc=0}
	dc=d\Omega=0\,.
\end{equation}
The condition $dc=0$ is solved by taking $c$ to be constant. The 2-form field $\Omega$ turns out to be unphysical, as well as the functions $m$ and $e$, which do not appear in any of the algebraic or differential conditions. Indeed, they can be removed by gauge transformations. To see this, consider a generalised vector of the form
\begin{equation}
	\Lambda=(\alpha\wedge vol_{S^1}+\Delta_1)+\Delta_{2}\wedge vol_{S^2}\wedge vol_{S^1}\,,
\end{equation}
with arbitrary 1-form $\alpha$ and 2-forms $\Delta_1$ and $\Delta_2$. This generates gauge transformations of the background and acting with it on the structures we find
\begin{equation}
	\begin{split}
		\mathcal{L}_\Lambda J_A&=\left(\frac{5^{1/6}\sqrt{2}}{R}\, d\Delta_2-d\alpha\wedge\chi-p\,d\Delta_1\right)\wedge dY_A\wedge vol_{S^1}\,,\\
		\mathcal{L}_\Lambda J_i&=0\,,\\
		\mathcal{L}_\Lambda \hat{K}&=-\frac{5^{1/6}\sqrt{2}}{R}\,d\alpha+\left(-c\,d\alpha\wedge vol_{S^2}\wedge vol_{S^1}+\left(\frac{5^{1/6}\sqrt{2}}{R}\,d\Delta_2-c\, d\Delta_1\right)\wedge vol_{S^2}\right)\,.
	\end{split}
\end{equation}
Since the generalised Lie derivative acts infinitesimally, we are only able to remove the exact parts of the differential forms. However, the parts which are closed but not exact will only appear in the gauge potentials of the background geometry and not in the field strengths. Therefore, these do not play a role in the equations of motion and thus in the construction of the background solution or its consistent truncations and we can therefore set them to zero, as we will henceforth do. By comparing with \eqref{eq:AdS5structures-1}, we see that with an appropriate choice of $\alpha$, $\Delta_1$ and $\Delta_2$ we can set $\Omega=e=m=0$, as we will take from now on. Similarly, we can and will set $c = 0$ as it will only appear in the gauge potentials and not the field strengths.

Taking this into account as well as the constraints \eqref{eq:diff-cond-vacua}, the general expression for the generalised structure of half-supersymmetric AdS$_5$ vacua reads
\begin{equation}\label{eq:structures-final}
	\begin{split}
		J_A&=\frac{5^{1/6}\sqrt{2}}{R}\,v_A+\Bigl( p\,dY_A\wedge vol_{S^1}+\chi\wedge dY_A+Y_A\, dp\wedge vol_{S^1}-Y_A\,d\chi\Bigr)\\
		&\quad +\frac{R}{5^{1/6}\sqrt{2}} \,\Bigl(-Y_A\,(p\,d\chi+\chi\wedge dp)\wedge vol_{S^2}\wedge vol_{S^1}+Y_A\,\chi\wedge d\chi\wedge vol_{S^2}-dp\wedge d\chi\wedge \theta_A\wedge vol_{S^1}\Bigr) \,, \\
		J_i&=d\Bigl(w_i\,\nu_1-\epsilon w_i\,\nu_2\Bigr)\\
		&\quad +\frac{R}{5^{1/6}\sqrt{2}}\, d \Bigl( \epsilon w_i(p\,d\nu_1-\chi\wedge \nu_2)\wedge vol_{S^2} + w_i(p\,d\nu_2+\chi\wedge \nu_1)\wedge vol_{S^2}\wedge vol_{S^1}\Bigl) \,, \\
		\hat{K}&=\frac{5^{1/6}\sqrt{2}}{R}\, v_{S^1}\,,
	\end{split}
\end{equation}
with $p$, $\nu_1$, $\nu_2$ and $\chi$ related by 
\begin{equation}\label{eq:M-theory-conditions-FINAL}
	\begin{split}
		d\nu_2\wedge dp&= -\nu_1\wedge d\chi\,,\qquad d\nu_1\wedge dp=\nu_2\wedge d\chi\,,\qquad \nu_1\wedge d\nu_2=dp\wedge d\chi \ge0\,,\\
		\nu_2\wedge d\nu_2&=\nu_1\wedge d\nu_1\,,\qquad\nu_1\wedge d\nu_2=-\nu_2\wedge d\nu_1\,,
	\end{split}
\end{equation}
which implies that 
\begin{equation}\label{eq:linear-idep-cond}
	\nu_1\wedge\nu_2\wedge dp\neq 0\,,
\end{equation}
everywhere except for the boundaries, as discussed above \eqref{eq:linear-idep-cond-1}.  The field $K=\frac{1}{5}J_u\wedge J^u$ is
\begin{equation}
	\begin{split}
		K &= \sqrt{\frac{2}{5}}\Bigl(\, -dp\wedge d\chi\wedge vol_{S^1} +\frac{R}{5^{1/6}\sqrt{2}} (p\, vol_{S^1}-\chi)\otimes dp \wedge d\chi\wedge vol_{S^2}\wedge vol_{S^1}\Bigr)\,,
	\end{split}
\end{equation}
and satisfies
\begin{equation} \label{eq:pGood}
	K\wedge\hK=\sqrt{\frac{2}{5}}\,p\,dp\wedge d\chi\wedge vol_{S^1}\wedge vol_{S^2} > 0 \,.
\end{equation}

\subsection{AdS$_5$ vacua fields}
Given the half-maximal structures \eqref{eq:structures-final}, we can construct the corresponding generalised metric $\gM_{MN}$ using \eqref{eq:Gen-Metric-from-structures}. From this, we obtain the corresponding 11-dimensional fields making use of the dictionary between supergravity objects and ExFT described in \eqref{eq:dictionary-E6}. The 11-dimensional metric, for instance, can be read from the components of the generalised metric with two five-form indices. For the structures \eqref{eq:structures-final}, and making use of the constraints \eqref{eq:M-theory-conditions-FINAL}, this part can be written as
\begin{equation}
	\begin{split}
		\gM_{\bar{m}\,\bar{n}}&=\frac{10^{1/3}}{R^2\,|dpd\chi|}\, \Bigl(|dpd\chi|\, |vol_{S^2}|\,p\Bigr)^{-4/3}\cdot \\
		&\quad\left[\frac{|dpd\chi|}{|dp\nu_1\nu_2|} \left(\,|dp\nu_1\nu_2| +2\,|dpd\chi| \,  p \right) (\nu_1\otimes\nu_1+\nu_2\otimes\nu _2+dp\otimes dp)
		\right. \\
		&\qquad +2\, |dp\nu_1\nu_2| \, p  \left(d\beta-\frac{|dpd\nu_1|}{|dp\nu_1\nu_2|}\nu_1-\frac{|dpd\nu_2|}{|dp\nu_1\nu_2|}\nu_2+\frac{|\nu_2 d\nu_2|}{|dp\nu_1\nu_2|}dp\right)^2	\\
		&\qquad\left.\vphantom{\frac{|dpd\chi|}{|dp\nu_1\nu_2|}}
		+|dpd\chi|\, p^2ds_{S^2}^2\right]_{m\,n}\,,
	\end{split}
\end{equation}
where $\beta$ is a local coordinate on $S^1$ and the quantities between $|\dots|$ are defined as $|dpd\chi|vol_3=dp\wedge d\chi$, and analogously for the rest. Similarly $vol_{S^2}=|vol_{S^2}|d\theta\wedge d\phi$ for some choice of local coordinates $\{\theta,\phi\}$ on $S^2$. The bars on top of the indices of the generalised metric in the left hand side of the expression indicate that they are actually five form indices, which have been contracted with epsilon symbols. 

In order to simplify this expression further, we will use diffeomorphisms on the 3-dimensional manifold $M_3$ to fix a certain coordinate frame $(x_1,x_2,y)$ and eliminate equivalent configurations. First, we define the coordinate $y$ to be the function
\begin{equation}
	y=\frac{p}{2 \times 5^{1/6} \sqrt{2}\,R^2} \,,
\end{equation}
where the rescaling factor is there to match the conventions in the literature.  Note that, because of \eqref{eq:pGood}, this function does not vanish at any point in the bulk of $M_3$ and therefore defines a good coordinate. Next, we will use a redefinition of the coordinates $x_1$ and $x_2$ to find a frame where $\nu_1$ and $\nu_2$ have no component along $dy$, namely,
\begin{equation}
	\begin{split}
		\nu_1&=a_1\, dx_1+a_2\, dx_2\,,\\
		\nu_2&=b_1\, dx_1+b_2\, dx_2\,,
	\end{split}
\end{equation}
where $a_1$, $a_2$, $b_1$ and $b_2$ are functions of the three coordinates $(x_1,x_2,y)$. Such a frame can always be found given the fact that $\nu_1$, $\nu_2$ and $dp$ are linear independent, as follows from \eqref{eq:linear-idep-cond}. 

With these choices, let us go back to the constraints \eqref{eq:M-theory-conditions-FINAL}. The last two involve  only the fields $\nu_1$ and $\nu_2$ and, given our gauge choice, can without loss of generality be solved by
\begin{equation}\label{eq:nus-gauge-fixing-v0}
	\begin{split}
		\nu_1&=c_2\, b_2\, dx_1+a_2(dx_2+c_1\, dx_1)\,,\\
		\nu_2&=-c_2\, a_2\, dx_1+b_2(dx_2+c_1\, dx_1)\,,	
	\end{split}
\end{equation}
where $c_1=c_1(x_1,x_2)$ and $c_2=c_2(x_1,x_2)$ are arbitrary functions not depending on the coordinate $y$. With such forms, the metric on the 3-dimensional manifold $M_3$ reads
\begin{equation}
	(\nu_1\otimes\nu_1+\nu_2\otimes\nu_2+dp\otimes dp)=(a_2^2+b_2^2)\,h_{ij}\, dx^idx^j+ \frac{1}{8 \times 5^{1/3} \,R^4}\,\, dy^2\,,
\end{equation}
where $h_{ij}$, with $i=1,2$, is constructed only out of the functions $c_1$ and $c_2$ and therefore does not depend on the coordinate $y$. Then, we can perform a last redefinition of the coordinates $(x_1,x_2)$ to fix a conformal gauge in which $h_{ij}\propto \delta_{ij}$ and the corresponding proportionality factor can be absorbed by a redefinition of the fields $(a_2,b_2)$. These manipulations are actually equivalent to setting $c_2=1$ and $c_1=0$ in \eqref{eq:nus-gauge-fixing-v0}. Finally, redefining the fields $(a_2,b_2)=2 \times 5^{1/6} \sqrt{2}\,R^2\, e^{D/2}(\sin\Theta,\cos\Theta)$ we are left with the forms
\begin{equation}\label{eq:nus-dp-gauge-fixed}
	\begin{split}
		\nu_1&=2 \times 5^{1/6} \sqrt{2}\,R^2\,e^{D/2}\,(\cos\Theta\, dx_1+\sin\Theta\, dx_2) \,, \\
		\nu_2&=2 \times 5^{1/6} \sqrt{2}\,R^2\,e^{D/2}\,(-\sin\Theta\, dx_1+\cos\Theta\, dx_2) \,, \\
		dp&=2 \times 5^{1/6} \sqrt{2}\,R^2\, dy \,,
	\end{split}
\end{equation}
with $D$ and $\Theta$ two arbitrary functions of $M_3$. With them, we can address the remaining three conditions in \eqref{eq:M-theory-conditions-FINAL}, which can now be generically solved by taking
\begin{equation}\label{eq:dchi-gauge-fixed}
	d\chi= -5^{1/6} \sqrt{2}\,R^2\,\Bigl(\epsilon^{ij}\partial_i D\, dx_j \wedge dy +\partial_y e^D\, dx_1\wedge dx_2 +2\,d(\Theta\, dy)\Bigr)\,.
\end{equation}
The consistency condition $d^2\chi=0$ requires
\begin{equation}\label{eq:Toda-equation}
	\partial_1^2 D+\partial_2^2 D+\partial_y^2 e^D=0\,,
\end{equation}
which is the Toda equation \eqref{eq:Toda-equation-(intro)}. 

Using the fields \eqref{eq:nus-dp-gauge-fixed} and \eqref{eq:dchi-gauge-fixed}, the condition $K\wedge\hK>0$ becomes
\begin{equation}\label{eq:positivity-condition-gauge-fixed}
	-16\,e^D \,R^6\, y\, \partial_yD\,vol_3\wedge vol_{S^2}\wedge vol_{S^1}>0\,,
\end{equation}
which we solve by taking
\begin{equation}
	y>0\,,\qquad \partial_y D<0\,.
\end{equation}
Then, from the dictionary \eqref{eq:dictionary-E6},
\begin{equation}
	\det g=(\det\mathcal{M}_{\bar{m}\,\bar{n}})^{-1/3}= \frac{4\, R^{12}\,e^{2 \,D}\,|vol_{S^2}|^2\,  (\partial_y D)^3}{ \left( y\, \partial_y D-1 \right)}\,,
\end{equation} 
and the internal metric becomes
\begin{equation}
	\begin{split}
		ds^2&=(\det g)^{2/3}(\mathcal{M}_{\bar{m}\,\bar{n}})\\
		&=f_4\,(e^D\,dx_idx^i+dy^2)+f_3\, (d\beta+A_i dx^i-d\Theta)^2+f_2\, ds^2_{S^2}
	\end{split}
\end{equation}
with
\begin{equation}
	\begin{split}
		f_4&=\frac{R^2\, e^{-4\,\lambda}}{1-y^2\,e^{-6\lambda}}\,,\qquad
		f_3=4\,R^2\,e^{2\lambda}\,(1- y^2 e^{-6\lambda})\,,\qquad
		f_2=R^2\, y^2\,e^{-4\lambda}\,,\\
		A_i&=\frac{1}{2}\,\epsilon_{ij}\partial_j D\,,
	\end{split}
\end{equation}
where
\begin{equation}
	e^{-6\lambda}=
	-\frac{\partial_y D}{y\, (1- y\, \partial_y D)}\,.
\end{equation}
We observe that the function $\Theta$ can be completely removed from the geometry through a local shift of coordinate $\beta$, and we will therefore from now on set $\Theta=0$. The AdS$_5$ warp factor can also be read off from the structures as in \eqref{eq:Warp} and in our case is
\begin{equation}
	f_1= |\det g|^{-1/3}\kappa^2 =4\,e^{2\lambda}\,,
\end{equation}
and the full metric becomes
\begin{equation}\label{eq:LLM-metric}
	ds^2=f_1 ds^2_{AdS_5}+f_4\,(e^D\,dx_idx^i+dy^2)+f_3\, (d\beta+A_i dx^i)^2+f_2\, ds^2_{S^2}\,.
\end{equation}
Finally, from the second equation in \eqref{eq:dictionary-E6} and using the gauge-fixed forms \eqref{eq:nus-dp-gauge-fixed} we obtain the gauge field
\begin{equation}
	\begin{split}
		C_{(3)}&=R^3\left(4y^3 e^{-6\lambda} (d\beta+A_idx^i-d\Theta)
		-4\left(\frac{1}{2\times 5^{1/6}\sqrt{2}R^2}\chi-y\,d\Theta+y\, A_i dx^i\right)\right)\wedge vol_{S^2}\,,
	\end{split}
\end{equation}
with field strength
\begin{equation}
	F_{(4)}=dC_{(3)}=R^3\left(d\left(4\,y^3\,e^{-6\,\lambda}(d\beta+A_i dx^i-d\Theta)\right)+d\hat{B}\right)\wedge vol_{S^2}\,,
\end{equation}
with
\begin{equation}
	d\hat{B}=2\left(y^2\partial_y\left(\frac{1}{ y}\partial_y e^D\right)dx_1\wedge dx_2+y\, \partial_1\partial_yD\, dx_2\wedge dy-y\, \partial_2\partial_yD\, dx_1\wedge dy\right)\,,
\end{equation}
and we observe that the function $\Theta$ can be removed by the same local shift of $\beta$ we used to remove it from the metric. These geometries are those obtained in \cite{Lin:2004nb}.

\section{Minimal consistent truncation}\label{sec:minimal-truncation}
As shown in \cite{Malek:2017njj} and reviewed in section \ref{sec:minimal-truncation-review}, we can use the generalised $\USp{4}$ structure of the AdS$_5$ vacua to immediately construct a consistent truncation to 5-dimensional $\SU{2} \times \UO$ gauged supergravity keeping only the gravitational supermultiplet. The truncation Ansatz is given by \eqref{eq:MinimalTruncation} and \eqref{eq:MinimalTruncationTensors} and all that is left to do is to use the dictionary between ExFT and 11-dimensional supergravity to compute the supergravity fields. Here we restrict ourselves for simplicity to the scalar sector, for which we find the uplift formulae
\begin{equation}
	\begin{split}
		ds^2&=\tilde f_4\,(e^D\,dx_idx^i+dy^2)+\tilde f_3\, (d\beta+A_i dx^i)^2+\tilde f_2\, ds^2_{S^2}\\
		F_{(4)}&=\left(d\left(4\,y^3\,e^{-6\,\tilde \lambda}(d\beta+A_i dx^i)\right)+d\hat{B}\right)\wedge vol_{S^2}\,,
	\end{split}
\end{equation}
where $A_idx^i$ and $d\hat{B}$ are the same forms as in the vacuum and 
\begin{equation}	
	\begin{split}
		\tilde f_4&=\frac{R^2\,X^{-2}\, e^{-4\,\tilde\lambda}}{1-y^2\,e^{-6\tilde\lambda}}\,,\qquad
		\tilde f_3=4\,R^2X^{4}\,e^{2\tilde\lambda}\,(1- y^2 e^{-6\tilde\lambda})\,,\qquad
		\tilde f_2=R^2 X^{-2}\, y^2\,e^{-4\tilde\lambda}\,,
	\end{split}
\end{equation}
with
\begin{equation}
	e^{-6\tilde\lambda}=
	-\frac{\partial_y D}{y\, (X^{-3}- y\, \partial_y D)}\,.
\end{equation}
This agrees with the consistent truncation constructed in \cite{Gauntlett:2007sm}.

\section{Consistent truncations with vector multiplets}\label{sec:vector-multiplets}
 In section \ref{sec:vector-multiplets-review}, we reviewed that for an AdS vacuum to admit a consistent truncation with vector multiplets around it, its internal manifold should admit $n$ extra vector multiplets $\bar{J}_{\bar{u}}$ satisfying the constraints \eqref{eq:vec-mult-algebraic-cond-rev} and  \eqref{eq:vec-mult-diff-cond-rev}. We will now use these conditions to classify all possible consistent truncations with vector multiplets around the half-maximal AdS$_5$ vacua. Using our results, it is straightforward to construct the consistent truncations with vector multiplets from the formulae \eqref{eq:VecTruncationAnsatz}, although we will not do so here.

In particular, \eqref{eq:vec-mult-diff-cond-rev} implies that the additional $\bar{J}_{\bar{u}}$ have to organise into representations of $(\SU{2}\times \UO)_R$. Furthermore, the explicit expressions of $J_i$ in \eqref{eq:structures-final} show that these are ``trivial'' generalised vector fields whose action via the generalised Lie derivative vanishes automatically
\begin{equation}
	\gL_{J_i} = 0 \,,
\end{equation}
for any tensor it acts on. This immediately implies that $f_{i\bar{u}}{}^{\bar{v}} = 0$, which is consistent with the outcome of the analysis of the embedding tensor constraints of the 5-dimensional gauged supergravity \cite{Louis:2015dca}. Moreover, the $\SU{2}$ representations of $\bar{J}_{\bar{u}}$ must be of odd dimensions, i.e. be representations of $\SO{3}$.

Together with the fact that $n\le 5$, we are then left with the following possibilities for the vector multiplets, which we discuss in the subsequent sections:
\begin{itemize}
	\item up to five $\SU{2}\times \UO$ singlets (sections \ref{s:Singlet} and \ref{s:Singlets}),
	\item one $\SU{2}$ triplet (section \ref{s:Triplet}),
	\item one $\SU{2}$ triplet and up to two singlets (section \ref{s:TripletSinglet}),
	\item one $\SU{2}$ quintuplet, i.e. the symmetric traceless representation of $\SO{3}$ (section \ref{s:Quintuplet}),
	\item up to two $\UO$ doublets (sections \ref{s:Doublet} and \ref{s:Doublets}),
	\item one $\SU{2}$ triplet and one $\UO$ doublet (section \ref{s:TripletDoublet}),
	\item one $\UO$ doublet and up to three singlets or two $\UO$ doublets and one singlet (section \ref{s:DoubletSinglets}).
\end{itemize}
By ``$\UO$ doublet'' we mean the complex $\UO$ representation with general charge $q$ under $\UO$. 
We will now analyse these various possibilities in turn and show that, at most, only the following possibilities are allowed:
\begin{itemize}
	\item up to three $\SU{2}\times \UO$ singlets,
	\item one $\SU{2}$ triplet,
	\item one $\UO$ doublet,
	\item one $\UO$ doublet and one singlet.
\end{itemize}

\subsection{One singlet under  $\Rsym$} \label{s:Singlet}
To construct a consistent truncation with one extra vector multiplet, the internal manifold should admit an extra generalised vector $\bar{J}$. The most general ansatz compatible with the symmetries is
\begin{equation}
	\begin{split}
		\bar J&=\bar t \,v_{S^1}+\Bigl(\bar p\,vol_{S^2}+\bar\Phi+\bar\nu\wedge vol_{S^1}\Bigr)
		+\Bigl(\bar\Pi\wedge vol_{S^2}\wedge vol_{S^1}+\bar z\, vol_3\wedge vol_{S^2}\Bigr)\,,
	\end{split}
\end{equation}
with $\bar t$, $\bar p$ and $\bar z$ arbitrary functions of $M_3$,  $\bar\nu$ a 1-form and $\bar\Phi$ and $\bar{\Pi}$ 2-forms. Using the structures \eqref{eq:structures-final}, the algebraic conditions \eqref{eq:vec-mult-algebraic-cond-rev} restrict the ansatz to
\begin{equation}
	\begin{split}
		\bar J&=\Bigl(\bar\Phi+\bar\nu\wedge vol_{S^1}\Bigr)
		+\frac{R}{5^{1/6}\sqrt{2}}\Bigl((p\,\bar\Phi-\chi\wedge\bar\nu)\wedge vol_{S^2}\wedge vol_{S^1}-\bar \Phi\wedge\chi\wedge vol_{S^2}\Bigr)\,,
	\end{split}
\end{equation}
where $\bar\nu$ and $\bar\Phi$ are a 1- and a 2-form satisfying the constraints
\begin{equation}\label{eq:algebraic-cond-vec-mult}
	\begin{split}
		\nu_2\wedge\bar\Phi&=-\bar\nu\wedge d\nu_1\,,\qquad \nu_1\wedge\bar\Phi=\bar\nu\wedge d\nu_2\,,\\
		dp\wedge\bar\Phi&=\bar\nu\wedge d\chi\,,\qquad \quad \bar\nu\wedge\bar\Phi=dp \wedge d\chi\,.
	\end{split}
\end{equation}
These constraints can be completely solved as follows. Given that $\nu_1$, $\nu_2$ and $\nu_3$ form a basis of 1-forms at each point of the manifold, one can write
\begin{equation}
	\bar\nu=v_1\,\nu_1+v_2\,\nu_2+v_3\,dp\,,
\end{equation}
with $v_1$, $v_2$ and $v_3$ three functions of $M_3$, and use the first three equations of \eqref{eq:algebraic-cond-vec-mult} to completely determine $\bar{\Phi}$ in terms of $\bar{\nu}$. In particular, using \eqref{eq:nus-dp-gauge-fixed} (with $\Theta=0$),
\begin{equation}
	\begin{split}
		\bar\Phi&=\frac{1}{2\times 5^{1/6}\sqrt{2}\, R^2}\left[\left(v_1 \partial_y D-e^{-\frac{1}{2} D}\, v_3\, \partial_1 D\right)dp\wedge \nu_2  
		+ \left(-v_2 \partial_y D +e^{-\frac{1}{2} D}v_3 \partial_2 D\right) dp\wedge\nu_1\right]\\
		&\quad +v_2\,d\nu_1-v_1\, d\nu_2 + v_3 \, d\chi\,,
	\end{split}
\end{equation}
and the last condition of \eqref{eq:algebraic-cond-vec-mult} becomes
\begin{equation}
	(v_1^2+v_2^2+v_3^2)\,dp\wedge d\chi=dp\wedge d\chi\,.
\end{equation}
Therefore, given the fact that $dp\wedge d\chi\neq 0$, a general solution to \eqref{eq:algebraic-cond-vec-mult} given the fields \eqref{eq:nus-dp-gauge-fixed} is
\begin{equation}\label{eq:alg-cond-vec-mult-singlet-SOL}
	\begin{split}
		\bar\nu&=v_1\,\nu_1+v_2\,\nu_2+v_3\,dp\,,\qquad \text{with}\quad v_1^2+v_2^2+v_3^2=1\,,\\
		\bar\Phi&=\frac{1}{2\times 5^{1/6}\sqrt{2}\, R^2}\left[\left(v_1 \partial_y D-e^{-\frac{1}{2} D}\, v_3\, \partial_1 D\right)dp\wedge \nu_2  
		+ \left(-v_2 \partial_y D +e^{-\frac{1}{2} D}v_3 \partial_2 D\right) dp\wedge\nu_1\right]\\
		&\quad +v_2\,d\nu_1-v_1\, d\nu_2 + v_3 \, d\chi\,.
	\end{split}
\end{equation}
Finally, the differential conditions \eqref{eq:vec-mult-diff-cond-rev} imply that
\begin{equation}
	\begin{split}
		&d\bar\nu=d\bar\Phi=0 \,.
	\end{split}
\end{equation}
In terms of the solution \eqref{eq:alg-cond-vec-mult-singlet-SOL}, the condition $d\bar{\nu}=0$ becomes
\begin{equation}\label{eq:diff-cond-SINGLET-VM-1}
	\begin{split}
		\partial_2\left(e^{\frac{1}{2}D}v_1\right)&=\partial_1\left(e^{\frac{1}{2}D}v_2\right)\,,\\
		\partial_1 v_3 &=\partial_y\left(e^{\frac{1}{2}D}v_1\right)\,,\\
		\partial_2 v_3 &=\partial_y\left(e^{\frac{1}{2}D}v_2\right)\,,
	\end{split}
\end{equation}
and using these, the condition $d\bar{\Phi}=0$ can be written as
\begin{equation}\label{eq:diff-cond-SINGLET-VM-2}
	\partial_1\left(e^{\frac{1}{2}D} (\partial_y D)^2\, v_1\right)
	+\partial_2\left(e^{\frac{1}{2}D} (\partial_y D)^2\, v_2\right)
	+e^{-D}\,\partial_y\left(e^{2\,D} (\partial_y D)^2\, v_3\right)=0\,.
\end{equation}

Thus, a consistent truncation with a single vector multiplet requires the existence of a one-form on $M_3$ such that its components in the $\nu_1$, $\nu_2$, $dp$ basis form a unit-norm triplet $\left(v_1,v_2,v_3\right)$ such that \eqref{eq:diff-cond-SINGLET-VM-1} and \eqref{eq:diff-cond-SINGLET-VM-2} are satisfied. The resulting 5-dimensional gauged supergravity has embedding tensor $f_{abc}$, $\xi_{ab}$ with $a = (u, 6)$ and whose only non-zero components are
\begin{equation}
	f_{ABC} = \coeff \epsilon_{ABC} \,, \qquad \xi_{ij} = \coeff \epsilon_{ij} \,.
\end{equation}
Hence the five-dimensional supergravity has $\SU{2} \times \UO$ gauging.

\subsection{Multiple $\Rsym$ singlets} \label{s:Singlets}
To analyse the more general case of $\Rsym$ singlets, we first consider the case of two $\Rsym$ singlets. Following the discussion in the previous section this is described by two generalised vectors of the form
\begin{equation}
	\begin{split}
		\bar J_{\bar{1}}&=\Bigl(\bar\Phi_1+\bar\nu_1\wedge vol_{S^1}\Bigr)
		+\frac{1}{\rho}\Bigl((p\,\bar\Phi_1-\chi\wedge\bar\nu_1)\wedge vol_{S^2}\wedge vol_{S^1}-\bar \Phi_1\wedge\chi\wedge vol_{S^2}\Bigr)\,,\\
		\bar J_{\bar{2}}&=\Bigl(\bar\Phi_2+\bar\nu_2\wedge vol_{S^1}\Bigr)
		+\frac{1}{\rho}\Bigl((p\,\bar\Phi_2-\chi\wedge\bar\nu_2)\wedge vol_{S^2}\wedge vol_{S^1}-\bar \Phi_2\wedge\chi\wedge vol_{S^2}\Bigr)\,,
	\end{split}
\end{equation}
where $\bar\nu_1=v_1\nu_1+v_2\nu_2+v_3 dp$, $\bar\nu_2=w_1\nu_1+w_2\nu_2+w_3 dp$ and the pairs $(\bar\nu_1,\bar\Phi_1)$ and  $(\bar\nu_2,\bar\Phi_2)$ individually satisfy \eqref{eq:alg-cond-vec-mult-singlet-SOL}, \eqref{eq:diff-cond-SINGLET-VM-1} and \eqref{eq:diff-cond-SINGLET-VM-2}. The only condition that remains to be checked is the algebraic constraint
\begin{equation}
	\bar{J}_{\bar{1}}\wedge\bar{J}_{\bar{2}}=0\,,
\end{equation}
which implies that
\begin{equation}\label{eq:alg-cond-vec-mult-2singlets}
	\bar\nu_1\wedge\bar\Phi_2+\bar\nu_2\wedge\bar\Phi_1=0\,.
\end{equation}
Using \eqref{eq:alg-cond-vec-mult-singlet-SOL}, this condition becomes
\begin{equation}
	(v_1\,w_1+v_2\,w_2+v_3\,w_3)=0\,,
\end{equation}
implying that $(v_1,v_2,v_3)$ and $(w_1,w_2,w_3)$ are perpendicular.

Generalising to $n\le 5$ vector multiplets transforming as singlets is straightforward: the extra generalised vectors will be characterised by $n$ pairs $(\bar\nu_{\bar{u}},\bar\Phi_{\bar{u}})$, $\bar{u}=1,\dots n$, each of them satisfying individually \eqref{eq:alg-cond-vec-mult-singlet-SOL}, \eqref{eq:diff-cond-SINGLET-VM-1} and \eqref{eq:diff-cond-SINGLET-VM-2}. Furthermore, the vectors formed with the components of the different $\bar\nu_{\bar{u}}$ in the $\{\nu_1,\nu_2,dp\}$ basis will have to be perpendicular pairwise. This condition implies that we can keep at most three $\Rsym$ singlets. The resulting gauged supergravity has embedding tensor $f_{abc}$, $\xi_{ab}$ with $a = (u, \bar{u})$, with $\bar{u} = 1, \ldots, n$ with $n \leq 3$ labelling the number of vector multiplets. The only non-zero components are
\begin{equation}
	f_{ABC} = \coeff \epsilon_{ABC} \,, \qquad \xi_{ij} = \coeff \epsilon_{ij} \,.
\end{equation}
Hence the five-dimensional supergravity has $\SU{2} \times \UO$ gauging.

\subsection{Triplet under $\SUt$} \label{s:Triplet}
We next analyse the possibility of having three vectors organising into a triplet under $\SUt$. The most general ansatz for the extra generalised vectors compatible with the symmetries is
\begin{equation}
	\begin{split}
		\bar J_{\bar A}&=\bar \rho \,v_A+\Bigl(\bar q\, y_A\,vol_{S^2}+\bar p\,dy_A\wedge vol_{S^1}+\bar f\,\theta_A\wedge vol_{S^1}+\bar\chi_1\wedge dy_A+\bar\chi_2\wedge\theta_A+y_A\,\bar\Phi+y_A\, \bar\nu\wedge vol_{S^1}\Bigr)\\
		&\quad+\Bigl(y_A\, \bar\Pi\wedge vol_{S^2}\wedge vol_{S^1}+\bar z\,y_A\, vol_3\wedge vol_{S^2}+\bar m\,  vol_3\wedge dy_A\wedge vol_{S^1}+\bar n\,  vol_3\wedge \theta_A\wedge vol_{S^1}\Bigr)\,.
	\end{split}
\end{equation}
After applying the constraints \eqref{eq:algebraic-cond-vec-mult} using the vacuum structures \eqref{eq:structures-final}, this reduces to
\begin{equation}\label{eq:vec-mult-triplet-gen-vec-final}
	\begin{split}
		\bar J_{\bar A}&= \rho \,v_A+\Bigl(p\,dy_A\wedge vol_{S^1}+\chi\wedge dy_A+y_A\,\bar\Phi+y_A\, \bar\nu\wedge vol_{S^1}\Bigr)\\
		&\quad+\frac{1}{\rho}\Bigl(y_A\, (p\,\bar\Phi-\chi\wedge\bar\nu)\wedge vol_{S^2}\wedge vol_{S^1}-y_A\, \bar \Phi\wedge\chi\wedge vol_{S^2}+d\chi\wedge dp\wedge \theta_A\wedge vol_{S^1}\Bigr)\,,
	\end{split}
\end{equation}
where $\bar\nu$ and $\bar\Phi$ satisfy the constraints
\begin{equation}
	\begin{split}
		\nu_2\wedge\bar\Phi&=-\bar\nu\wedge d\nu_1\,,\qquad \nu_1\wedge\bar\Phi=\bar\nu\wedge d\nu_2\,,\\
		dp\wedge\bar\Phi&=\bar\nu\wedge d\chi\,,\qquad\bar\nu\wedge\bar\Phi=dp \wedge d\chi\,.
	\end{split}
\end{equation}
These are precisely the constraints \eqref{eq:algebraic-cond-vec-mult} and are therefore again solved by \eqref{eq:alg-cond-vec-mult-singlet-SOL}. However, the differential conditions \eqref{eq:vec-mult-diff-cond-rev} now imply 
\begin{equation}
	\begin{split}
		\bar\nu&=dp\,,\\
		-p\, d\bar\Phi+2 \, \bar\nu\wedge\bar\Phi&=0\,.
	\end{split}
\end{equation}
The first of these conditions fixes $(v_1,v_2,v_3)=(0,0,1)$ and
\begin{equation}
	\begin{split}
		\bar\Phi&=d\chi+\frac{1}{2\times 5^{1/6}\sqrt{2}\, R^2}\,e^{-\frac{1}{2} D}\, (-\partial_1 D\,dp\wedge \nu_2  + \partial_2 D\, dp\wedge\nu_1) \\
		&=d\chi+2\times 5^{1/6}\sqrt{2}\, R^2\,\epsilon_{ij}\,\partial_i D\, dx^j\wedge dy\,.
	\end{split}
\end{equation}
Then, taking $D$ to be a solution to the Toda equation \eqref{eq:Toda-equation}, the second condition is satisfied only when 
\begin{equation}
	y\, \partial_y^2 e^D-\partial_y e^D=0\,,
\end{equation}
which is in general solved by
\begin{equation}
	e^D=f_1(x_1,x_2)\,y^2+f_2(x_1,x_2)\,,
\end{equation}
with  $f_1(x_1,x_2)$ and $f_2(x_1,x_2)$ arbitrary functions of $(x_1,x_2)$. However, since $D$ is a solution to the Toda equation and $\partial_y D < 0$ and $y > 0$, the function $D$ is restricted to be
\begin{equation}\label{eq:vec-mult-triplet-SOL-for-D}
	e^D=e^{\sigma(x_1,x_2)}(- y^2+\text{constant})\,,
\end{equation}
with $\sigma$ any function of $(x_1,x_2)$ satisfying
\begin{equation}
	\partial_1^2\sigma + \partial_2^2 \sigma - 2 e^\sigma = 0 \,.
\end{equation}
Therefore, only in the cases where the vacuum is characterised by a function $D$ of the form \eqref{eq:vec-mult-triplet-SOL-for-D}, the truncation keeping three vector multiplets transforming as a triplet under $\SUt_R$ is consistent. This form implies that the internal space is $S^4$ fibred over a Riemann surface as in \cite{Maldacena:2000mw}, as we will discuss in more detail in section \ref{sec:AdS-SL(5)}.

Once all constraints are fixed, the generalised vectors \eqref{eq:vec-mult-triplet-gen-vec-final} satisfy 
\begin{equation}
	\begin{split}
		\mathcal{L}_{J_A}\bar J_{\bar B}&=-\frac{5^{1/6}\sqrt{2}}{R}\,\epsilon_{A\bar B\bar C}\,\bar J_{\bar C} \,, \\
		\mathcal{L}_{\bar J_{\bar A}} J_{ B}&=-\frac{5^{1/6}\sqrt{2}}{R}\,\epsilon_{\bar A B\bar C}\,\bar J_{\bar C} \,, \\
		\mathcal{L}_{\bar J_{\bar A}}\bar J_{\bar B}&=\frac{5^{1/6}\sqrt{2}}{R}\,\epsilon_{\bar A\bar B C}\, J_{C}-\frac{2\times 5^{1/6}\sqrt{2}}{R}\,\epsilon_{\bar A \bar B\bar C}\,\bar J_{\bar C} \,,
	\end{split}
\end{equation}
which implies that the embedding tensor $f_{abc}$, $\xi_{ab}$ with $a = (u, \bar{A})$ is given by
\begin{equation}
	\begin{split}
	f_{ABC} = \coeff \epsilon_{ABC} \,, \qquad f_{\bar{A}\bar{B}\bar{C}} &= - 2\, \coeff \epsilon_{ABC} \,, \qquad f_{A\bar{B}\bar{C}} = - \coeff \epsilon_{ABC} \,, \\
	&\xi_{ij} = \coeff \epsilon_{ij} \,,
	\end{split}
\end{equation}
and all other components vanishing. Therefore, the gauge group of the resulting supergravity is ISO(3)$\times$U(1). Explicit uplift expressions for this consistent truncation have been constructed using SO(5) maximal gauged supergravity in \cite{Cheung:2019pge}, while this consistent truncation has also been studied using generalised geometry in \cite{Cassani:2019vcl}. As we've shown here, this consistent truncation exists if and only if the AdS$_5$ vacuum is of the type of $S^4$ fibred over a Riemann surface.

\subsection{One $\SUt_R$ triplet and multiple $\Rsym$ singlets} \label{s:TripletSinglet}
For a background to allow a truncation keeping one triplet of $\SUt$ and multiple singlets, it has to allow both of them independently, satisfying the conditions described in the previous sections, together with extra compatibility conditions. However, in the following we will see that not even the case with one singlet is possible. Let us consider the generalised vectors $\bar J_{\bar{A}}$ and $\bar{J}$ describing  the triplet and the singlet respectively. The algebraic compatibility condition that needs to be satisfied is
\begin{equation}
	\bar{J}_{\bar{A}}\wedge \bar{J}=0 \,.
\end{equation} 
This condition implies
\begin{equation}
	\bar\nu^{(\text{sing.})}\wedge\bar\Phi^{(\text{trip.})}+
	\bar\nu^{(\text{trip.})}\wedge\bar\Phi^{(\text{sing.})}=0\,,
\end{equation}
which, taking $\bar\nu^{(\text{sing.})}=v_1\,\nu_1+v_2\,\nu_2+v_3\,dp$ and $	\bar\nu^{(\text{trip.})}=w_1\,\nu_1+w_2\,\nu_2+w_3\,dp$, is solved by $(v_1,v_2,v_3)$ and $(w_1,w_2,w_3)$ being perpendicular, as discussed below \eqref{eq:alg-cond-vec-mult-2singlets}. But $\bar\nu^{(\text{trip.})}=dp$, implying that $\bar\nu^{(\text{sing.})}=v_1\,\nu_1+v_2\,\nu_2$ in which case \eqref{eq:diff-cond-SINGLET-VM-1} becomes
\begin{equation}
	\begin{split}
		\partial_y\left(e^{\frac{1}{2}D}v_1\right)
		&=\partial_y\left(e^{\frac{1}{2}D}v_2\right)=0\,,\\
		\partial_2\left(e^{\frac{1}{2}D}v_1\right)
		&=\partial_1\left(e^{\frac{1}{2}D}v_2\right)\,.
	\end{split}
\end{equation}
The first equations are solved by taking
\begin{equation}
	(v_1,v_2,0)=e^{-\frac{1}{2}D} (V_1(x_1,x_2),V_2(x_1,x_2),0)\,,
\end{equation}
in which case, taking $y$-derivatives on both sides of the condition
\begin{equation}
	1=v_1^2+v_2^2+v_3^2=e^{-D}(V_1^2+V_2^2)\,,
\end{equation}
implies that $\partial_y D=0$. However, this is not an acceptable condition on $D$, since it would contradict the condition \eqref{eq:positivity-condition-gauge-fixed}. Therefore, a truncation keeping an $\SUt_R$ triplet and a singlet is not possible, nor is a truncation keeping a triplet and multiple singlets. 

\subsection{Quintuplet of $\SUt_R$} \label{s:Quintuplet}
Next we investigate the possibility of adding five extra $\bar J_{(\!(AB)\!)}$ transforming in the symmetric traceless representation of $\SUt$. The most general ansatz is
\begin{equation}
	\begin{split}
		\bar J_{(\!(AB)\!)}&=\bar\rho\, y_{(\!(A} \,v_{B)\!)}\\
		&\quad+\Bigl(\bar q\, y_{(\!(AB)\!)}\,vol_{S^2}+\bar p\,y_{(\!(A}dy_{B)\!)}\wedge vol_{S^1}+\bar f\,y_{(\!(A}\theta_{B)\!)}\wedge vol_{S^1}\\
		&\qquad+y_{(\!(A|}\bar\chi_1\wedge dy_{|B)\!)}+y_{(\!(A|}\bar\chi_2\wedge\theta_{|B)\!)}+y_{(\!(AB)\!)}\,\bar\Phi+y_{(\!(AB)\!)}\, \bar\nu\wedge vol_{S^1}\Bigr)\\
		&\quad+\Bigl(y_{(\!(AB)\!)}\, \bar\Pi\wedge vol_{S^2}\wedge vol_{S^1}+\bar z\,y_{(\!(AB)\!)}\, vol_3\wedge vol_{S^2}+\bar m\, y_{(\!(A|} vol_3\wedge dy_{|B)\!)}\wedge vol_{S^1}\\
		&\qquad\quad+\bar n\, y_{(\!(A|} vol_3\wedge \theta_{|B)\!)}\wedge vol_{S^1}\Bigr)\,,
	\end{split}
\end{equation}
where $(\!(\dots)\!)$ indicate traceless symmetrisation. For instance,
\begin{equation}
	y_{(\!(AB)\!)}=y_A\,y_B-\frac{1}{3}\delta_{AB}\,.
\end{equation} 
The conditions $J_u\wedge \bar J_{(\!(AB)\!)}=0$ impose 
\begin{equation}
	\begin{split}
		\bar J_{(\!(AB)\!)}&=\bar\rho\, y_{(\!(A} \,v_{B)\!)}\\
		&\quad+\left(\frac{\bar\rho}{\rho}\, p\,y_{(\!(A}dy_{B)\!)}\wedge vol_{S^1}
		+\frac{\bar\rho}{\rho}\,y_{(\!(A|}\,\chi\wedge dy_{|B)\!)}+y_{(\!(AB)\!)}\,\bar\Phi+y_{(\!(AB)\!)}\, \bar\nu\wedge vol_{S^1}\right)\\
		&\quad+\left(\frac{1}{\rho}\,y_{(\!(AB)\!)}\, (p\,\bar\Phi-\chi\wedge\bar\nu)\wedge vol_{S^2}\wedge vol_{S^1}-\frac{1}{\rho}\,y_{(\!(AB)\!)}\, \bar \Phi\wedge\chi\wedge vol_{S^2}\right.\\
		&\qquad\quad \left. +\frac{\bar\rho}{\rho^2}\, y_{(\!(A|} \,d\chi\wedge dp\wedge \theta_{|B)\!)}\wedge vol_{S^1}\right) \,,
	\end{split}
\end{equation}
where $\bar\nu$ an $\bar\Phi$ satisfy
\begin{equation}
	\begin{split}
		\nu_2\wedge\bar\Phi&=-\bar\nu\wedge d\nu_1\,,\qquad \nu_1\wedge\bar\Phi=\bar\nu\wedge d\nu_2\,.
	\end{split}
\end{equation}
However, the condition $\bar J_{(\!(A_1 A_2)\!)}\wedge\bar J_{(\!(B_1 B_2)\!)}=-\delta_{(\!(A_1A_2)\!)(\!(B_1B_2)\!)}\,K$, with $\delta_{(\!(A_1A_2)\!)(\!(B_1B_2)\!)}$ the antisymmetric traceless version of delta, cannot be satisfied. Computing the 4-form part of  $\bar J_{(\!(A_1 A_2)\!)}\wedge\bar J_{(\!(B_1 B_2)\!)}$ we find
\begin{equation}
	\begin{split}
		&(\bar J_{(\!(A_1 A_2)\!)}\wedge\bar J_{(\!(B_1 B_2)\!)})_{(4)}=\\
		&\sqrt{\frac{2}{5}}\left(\frac{\bar\rho^2}{2\,\rho ^2}\, [\iota_{y_{(\!(A_1}\!v_{A_2)\!)}}y_{(\!(B_1}\theta_{B_2)\!)}+\iota_{y_{(\!(B_1}\!v_{B_2)\!)}}y_{(\!(A_1}\theta_{A_2)\!)}]\, dp\wedge d\chi+ y_{(\!(A_1A_2)\!)}y_{(\!(B_1B_2)\!)}\,\bar\nu\wedge\bar\Phi \right)\wedge vol_{S^1}\,.
	\end{split}
\end{equation}
However, $\iota_{y_{(\!(A_1}\!v_{A_2)\!)}}y_{(\!(B_1}\theta_{B_2)\!)}$ is
\begin{equation}
	\begin{split}
		\iota_{y_{(\!(A_1}\!v_{A_2)\!)}}y_{(\!(B_1}\theta_{B_2)\!)}&=\frac{1}{4} (\delta _{A_2B_2}\, y_{(\!(A_1B_1)\!)}+\delta _{A_2B_1} y_{(\!(A_1B_2)\!)}+\delta_{A_1B_2} y_{(\!(A_2B_1)\!)}+\delta _{A_1B_1} y_{(\!(A_2B_2)\!)})\\
		&\quad-\frac{1}{3} (\delta _{B_1B_2}  y_{(\!(A_1A_2)\!)}+\delta _{A_1A_2} y_{(\!(B_1B_2)\!)})+\frac{1}{6} (\delta _{A_1B_2} \delta _{A_2B_1}+\delta _{A_1B_1} \delta _{A_2B_2})\\
		&\quad -\frac{1}{9} \delta _{A_1A_2} \delta _{B_1B_2}- y_{(\!(A_1A_2)\!)} y_{(\!(B_1B_2)\!)} \,,
	\end{split}
\end{equation}
which gives more terms than just $\delta$'s and $y_{(\!(A_1A_2)\!)} y_{(\!(B_1B_2)\!)}$. These extra terms do not cancel amongst themselves and therefore the only solution is that all of them vanish, making it impossible to construct a truncation with five vector multiplets transforming as a quintuplet of $\SUt_R$.

\subsection{Doublet of $\UO$} \label{s:Doublet}
We next investigate the possibility of constructing a truncation with vector multiplets transforming as a $\UO_R$ doublet of general charge $q$, which is not necessarily the same as that of the $\UO$ vacuum structures. In particular, we look for extra generalised vectors transforming as
\begin{equation}
	\gL_\hK\bar\J_{\bar{\imath}}=-\frac{5^{1/6}\sqrt{2}}{R}\, q\, \epsilon_{\bar\imath\bar\jmath}\,\bar J^{\bar\jmath} \,,
\end{equation}
with $q\in\mathbb{Z}$ the $\UO$ charge. Using the functions $w_{(q)i}$ and differential forms $dw_{(q)i}$ defined in appendix \ref{sec:app-spheres}, the most general ansatz for $\bar{J}_{\bar\imath}$ is  
\begin{equation}
	\begin{split}
		\bar J_{\bar\imath}&=\Bigl(\bar b_1\,w_{(q)i}\,vol_{S^2}+\bar b_2\,\epsilon w_{(q)i}\,vol_{S^2}+w_{(q)i}\,\bar\Phi_1+\epsilon w_{(q)i}\,\bar\Phi_2+w_i\,\bar\nu_1\wedge vol_{S^1}+\epsilon w_{(q)i}\,\bar\nu_2\wedge vol_{S^1}\Bigr)\\
		&\quad+\Bigl( w_{(q)i}\,\bar\Pi_1\wedge vol_{S^2}\wedge vol_{S^1}+\epsilon w_{(q)i}\,\bar\Pi_2\wedge vol_{S^2}\wedge vol_{S^1}\\
		&\qquad\qquad+\bar z_1\, w_{(q)i}\, vol_3\wedge vol_{S^2}+\bar z_2\,\epsilon w_{(q)i}\, vol_3\wedge vol_{S^2}\Bigr)\,.
	\end{split}
\end{equation}
After applying the algebraic constraints \eqref{eq:vec-mult-algebraic-cond-rev}, this becomes
\begin{equation}
	\begin{split}
		\bar J_{\bar\imath}&=\Bigl(w_{(q)i}\,\bar\Phi_2+\epsilon w_{(q)i}\,\bar\Phi_1+w_i\,\bar\nu_2\wedge vol_{S^1}+\epsilon w_{(q)i}\,\bar\nu_1\wedge vol_{S^1}\Bigr)\\
		&\quad+\frac{1}{\rho}\Bigl( w_{(q)i}\,(p\,\bar\Phi_2-\chi\wedge\bar\nu_2)\wedge vol_{S^2}\wedge vol_{S^1}+\epsilon w_{(q)i}\,(p\,\bar\Phi_1-\chi\wedge\bar\nu_1)\wedge vol_{S^2}\wedge vol_{S^1}\\
		&\qquad\qquad- w_{(q)i}\, \bar\Phi_2\wedge\chi\wedge vol_{S^2}-\epsilon w_{(q)i}\, \bar\Phi_1\wedge\chi\wedge vol_{S^2}\Bigr)\,,
	\end{split}
\end{equation}
where $\bar\nu_i$ and $\bar\Phi_i$ satisfy the constraints
\begin{equation}\label{eq:alg-cond-doublet-VM-1}
	\begin{split}
		\nu_2\wedge\bar\Phi_i&=-\bar\nu_i\wedge d\nu_1\,,\qquad \nu_1\wedge\bar\Phi_i=\bar\nu_i\wedge d\nu_2\,,\\
		dp\wedge\bar\Phi_i&=\bar\nu_i\wedge d\chi\,,\qquad\quad \bar\nu_i\wedge\bar\Phi_i=dp \wedge d\chi\,,
	\end{split}
\end{equation}
for $i=1,2$, as well as
\begin{equation}\label{eq:alg-cond-doublet-VM-2}
	\bar\nu_2\wedge\bar\Phi_1+\bar\nu_1\wedge\bar\Phi_2=0\,.
\end{equation}
The conditions  \eqref{eq:alg-cond-doublet-VM-1} are solved by two pairs $(\bar\nu_1,\bar\Phi_1)$ and $(\bar\nu_2,\bar\Phi_2)$ each of them individually satisfying \eqref{eq:alg-cond-vec-mult-singlet-SOL}. Furthermore, taking $\bar\nu_1=v_1\,\nu_1+v_2\,\nu_2+v_3\,dp$ and $\bar\nu_1=w_1\,\nu_1+w_2\,\nu_2+w_3\,dp$, the condition \eqref{eq:alg-cond-doublet-VM-2} is solved by taking $(v_1,v_2,v_3)$ and $(w_1,w_2,w_3)$ to be perpendicular, as discussed below  \eqref{eq:alg-cond-vec-mult-2singlets}. Therefore, we must have
\begin{equation} \label{eq:alg-cond-Doublet-VM-1}
	\begin{split}
		v_1^2 + v_2^2 + v_3^2 = w_1^2 + w_2^2 + w_3^2 &= 1 \,, \\
		v_1\, w_1 + v_2\, w_2 + v_3\, w_3 &=0 \,.
	\end{split}
\end{equation}

The differential constraints \eqref{eq:vec-mult-diff-cond-rev} imply
\begin{equation}
	\bar\Phi_1=-\frac{1}{q}\,d\bar\nu_2\,,\qquad \bar\Phi_2=\frac{1}{q}\,d\bar\nu_1\,,
\end{equation}
which in terms of  $(v_1,v_2,v_3)$ and $(w_1,w_2,w_3)$ and using \eqref{eq:alg-cond-vec-mult-singlet-SOL} become
\begin{equation}\label{eq:diff-cond-Doublet-VM-1}
	\begin{split}
		q \left(v_1\, \partial_1 D +v_2\, \partial_2 D+2\,v_3\, \partial_y e^{\frac{1}{2} D}\right)&=2\, e^{-\frac{1}{2}D} \left(\partial_1\left(e^{\frac{1}{2}D} w_2\right)-\partial_2\left(e^{\frac{1}{2}D} w_1\right)\right)\,\\
		q \left(2\,v_2\, \partial_y e^{\frac{1}{2}D}-v_3\,\partial_2 D \right)&=2 \left(\partial_y\left(e^{\frac{1}{2}D}w_1\right)-\partial_1 w_3 \right)\,,\\		q \left(2\,v_1\, \partial_y e^{\frac{1}{2}D}-v_3\,\partial_1 D \right)&=-2 \left(\partial_y\left(e^{\frac{1}{2}D}w_2\right)-\partial_2 w_3 \right)\,,
	\end{split}
\end{equation}
and
\begin{equation}\label{eq:diff-cond-Doublet-VM-2}
	\begin{split}
		q \left(w_1\, \partial_1 D +w_2\, \partial_2 D+2\,w_3\, \partial_y e^{\frac{1}{2} D}\right)&=-2\, e^{-\frac{1}{2}D} \left(\partial_1\left(e^{\frac{1}{2}D} v_2\right)-\partial_2\left(e^{\frac{1}{2}D} v_1\right)\right)\,\\
		q \left(2\,w_2\, \partial_y e^{\frac{1}{2}D}-w_3\,\partial_2 D \right)&=-2 \left(\partial_y\left(e^{\frac{1}{2}D}v_1\right)-\partial_1 v_3 \right)\,,\\		q \left(2\,w_1\, \partial_y e^{\frac{1}{2}D}-w_3\,\partial_1 D \right)&=2 \left(\partial_y\left(e^{\frac{1}{2}D}v_2\right)-\partial_2 v_3 \right)\,.
	\end{split}
\end{equation}
With these conditions satisfied, we find
\begin{equation}
	\begin{split}
		\mathcal{L}_{J_A}\bar J_{\bar\imath}&=\mathcal{L}_{\bar J_{\bar\imath}} J_A=\mathcal{L}_{J_i}\bar J_{\bar\imath}=\mathcal{L}_{\bar J_{\bar\imath}} J_i=\mathcal{L}_{\bar J_{\bar\imath}} \bar J_{\bar\jmath}=0\,.
	\end{split}
\end{equation}
Thus, the lower-dimensional gauged supergravity has embedding tensor $f_{abc}$, $\xi_{ab}$, with $a = (u, \bar{i})$, whose only non-zero components are given by
\begin{equation}
	f_{ABC} = \coeff \epsilon_{ABC} \,, \qquad \xi_{ij} = \coeff \epsilon_{ij} \,, \qquad \xi_{\bar{i}\bar{j}} = -\coeff \epsilon_{ij}  \,.
\end{equation}
The resulting gauge group is still given by $\SU{2} \times \UO$ but  the $\xi_{\bar{i}\bar{j}}$ term changes the embedding of the $\UO$ inside $\SO{5,2}$ so that it is a linear combination of $\UO \subset \SO{5} \subset \SO{5,2}$ and $\UO \simeq \SO{2} \subset \SO{5,2}$, see \eqref{eq:gauging}.

This consistent truncation can be used to uplift the RG flows of \cite{Bobev:2018sgr} between ${\cal N}=4$ and ${\cal N}=2$ AdS$_5$ vacua. It would therefore be particularly interesting to find solutions of the conditions \eqref{eq:alg-cond-Doublet-VM-1}, \eqref{eq:diff-cond-Doublet-VM-1}, \eqref{eq:diff-cond-Doublet-VM-2}.

\subsection{Two $\UO_R$ doublets} \label{s:Doublets}
Following the results of the previous section, we consider two doublets of generalised vectors $\bar{J}^{(1)}_{\bar{\imath}}$ and $\bar{J}^{(2)}_{\bar{\imath}}$ characterised by $(\bar\nu^{(1)}_i,\bar\Phi^{(1)}_i)$ and $(\bar\nu^{(2)}_i,\bar\Phi^{(2)}_i)$, with $i=1,2$, respectively. They have to satisfy the condition
\begin{equation}
	\bar J^{(1)}_{\bar\imath}\wedge\bar{J}^{(2)}_{\bar\jmath}=0\,,
\end{equation}
which implies
\begin{equation}
	\bar\nu^{(1)}_{i}\wedge\bar\Phi^{(2)}_j+\bar\nu^{(2)}_j\wedge\bar\Phi^{(1)}_i=0 \,,
\end{equation}
for $i,j=1,2$. Following the discussion  under \eqref{eq:alg-cond-vec-mult-2singlets}, these equations are solved by taking the vectors formed by the components of $\bar\nu^{(1)}_i$ and $\bar\nu^{(2)}_i$ in the $(\nu_1,\nu_2,dp)$ basis to be perpendicular pairwise.  However, the vectors associated to the two $\bar\nu^{(1)}_i$ (and the same for $\bar\nu^{(2)}_i$) are already perpendicular among themselves, as discussed in the previous section, which makes the algebraic constraints impossible to solve. Therefore, it is not possible to have these consistent truncations.

\subsection{One $\SUt_R$ triplet and one $\UO_R$ doublet} \label{s:TripletDoublet}
The algebraic constraint
\begin{equation}
	\bar{J}_{\bar{A}}\wedge\bar{J}_{\bar\imath}=0\,,
\end{equation}
is again solved by 
\begin{equation}
	\bar\nu^{(\text{trip.})}\wedge\bar\Phi_i^{(\text{doub.})}+\bar\nu_i^{(\text{doub.})}\wedge\bar\Phi^{(\text{trip.})}=0\,,
\end{equation}
for $i=1,2$, implying that the vectors formed by the components of $\bar\nu_i^{(\text{doub.})}$ in the $(\nu_1,\nu_2,dp)$ basis have to be perpendicular to the vector characterising $\bar\nu^{(\text{trip.})}$. Given the fact that $\bar\nu^{(\text{trip.})}=dp$ and that the vectors characterising $\bar\nu_i^{(\text{doub.})}$ are also perpendicular among themselves and of norm one, the most general solution is $\bar\nu_1^{(\text{doub.})}=v_1\,\nu_1+v_2\,\nu_2$, with $v_1^2+v_2^2=1$, and either $\bar\nu_2^{(\text{doub.})}=v_2\,\nu_1-v_1\,\nu_2$ or $\bar\nu_2^{(\text{doub.})}=-v_2\,\nu_1+v_1\,\nu_2$. In the first case, the differential equations \eqref{eq:diff-cond-Doublet-VM-1} and \eqref{eq:diff-cond-Doublet-VM-2} reduce to 
\begin{equation}\label{eq:diff-cond-doub+trip-VM}
	\begin{split}
		(q-1)\,v_1\,\partial_y e^{\frac{1}{2}D}&=e^{\frac{1}{2}D}\partial_y v_1\,,\\
		(q-1)\,v_2\,\partial_y e^{\frac{1}{2}D}&=e^{\frac{1}{2}D}\partial_y v_2\,,\\
		(q+1)\,(v_1\,\partial_1 D+v_2\,\partial_2 D)&=-2\,(\partial_1 v_1+\partial_2 v_2)\,,\\
		(q+1)\,(v_2\,\partial_1 D-v_1\,\partial_2 D)&=-2\,(\partial_1 v_2-\partial_2 v_1)\,.
	\end{split}
\end{equation}
The first two equations are solved by
\begin{equation}
	v_1=V_1(x_1,x_2)\,e^{\frac{q-1}{2}D}\,,\qquad v_2=V_2(x_1,x_2)\,e^{\frac{q-1}{2}D}\,,
\end{equation}
in which case, given the fact that $\partial_y D\neq 0$ (otherwise $dp\wedge d\chi=0$), the condition $v_1^2+v_2^2=1$ can only be solved for the case $q=1$. In this case, $\partial_y v_1=\partial_y v_2=0$ and the last two equations in \eqref{eq:diff-cond-doub+trip-VM} can only have solutions if the $y$-dependence in $D$ factorises, namely
\begin{equation}
	e^D=f(y)\,e^{\sigma (x_1,x_2)}\,,
\end{equation}
which, like in section \ref{s:Triplet}, implies that the internal space is a $S^4$ fibration over a Riemann surface, as we also discuss in more detail in section \ref{sec:AdS-SL(5)}. Using the fact that $V_1^2+V_2^2=1$ the last two equations of \eqref{eq:diff-cond-doub+trip-VM} become
\begin{equation}
	\begin{split}
		\partial_1 \sigma&=V_2\,\partial_2 V_1-V_1 \partial_2 V_2 =\partial_2 \Bigl(\arcsin(V_1)\Bigr)\,,\\
		\partial_2 \sigma&=V_1\,\partial_1 V_2-V_2 \partial_1 V_1 =-\partial_1 \Bigl(\arcsin(V_1)\Bigr)\,,
	\end{split}
\end{equation}
which are Cauchy-Riemann-type equations. However, if these equations had a solution for $V_1$, then the function $\sigma$ would be harmonic, which is not compatible with $D$ being a solution to the Toda equation with $\partial_y D\neq 0$.

Considering the case where $\bar\nu_2=^{\text{(doub.)}}-v_1\,\nu_1+v_2\,\nu_2$ one reaches an analogous result and we therefore conclude that having a truncation with a $\UO_R$ doublet and a $\SUt_R$ triplet is not possible.

\subsection{Multiple $\UO_R$ doublets with multiple $\Rsym$ singlets} \label{s:DoubletSinglets}
We have already seen in section \ref{s:Doublets}, that it is impossible to keep multiple $\UO_R$ doublets. Therefore, the case of multiple $\UO_R$ doublets with some $\Rsym$ singlets is also not possible.

This leaves us with the case of one $\UO_R$ doublets together with multiple $\Rsym$ singlets. To analyse this, we will begin with the case of one $\UO_R$ doublet and one $\Rsym$ singlet. As in the previous cases, the algebraic constraint
\begin{equation}
	\bar{J}\wedge\bar{J}_{\bar\imath}=0\,,
\end{equation}
implies that  
\begin{equation}
	\bar\nu^{(\text{sing.})}\wedge\bar\Phi_i^{(\text{doub.})}+\bar\nu_i^{(\text{doub.})}\wedge\bar\Phi^{(\text{sing.})}=0\,,
\end{equation}
for $i=1,2$, which is again solved by taking the vectors formed by the components of $\bar\nu_i^{(\text{doub.})}$ in the $(\nu_1,\nu_2,dp)$ basis have to be perpendicular to the vector characterising $\bar\nu^{(\text{sing.})}$.

If we want to keep multiple $\Rsym$ singlets, each of them would have to contain such a 3-vector perpendicular to the two 3-vectors of the $\UO_R$ doublet as well as the other $\Rsym$ singlets. This implies that we can keep at most one singlet together with one $\UO_R$ doublet. The resulting embedding tensor $f_{abc}$, $\xi_{ab}$, with $a = (u, \bar{i}, 8)$, now has as its only non-zero components
\begin{equation}
	f_{ABC} = \coeff \epsilon_{ABC} \,, \qquad \xi_{ij} = \coeff \epsilon_{ij} \,, \qquad \xi_{\bar{i}\bar{j}} = -\coeff q\, \epsilon_{ij}  \,.
\end{equation}
The gauging is again given by $\SU{2} \times \UO$ but due to $\xi_{\bar{i}\bar{j}} \neq 0$ the embedding of the $\UO$ inside $\SO{5,3}$ is again a linear combination of $\UO \subset \SO{5} \subset \SO{5,3}$ and $\UO \subset \SO{3} \subset \SO{5,3}$, see \eqref{eq:gauging}.

\section{AdS$_5$ vacua of maximal 7d gauged supergravity via SL(5) ExFT}\label{sec:AdS-SL(5)}
An important class of half-maximal AdS$_5$ vacua arise as near horizon limit of M5 branes wrapped on Riemann surfaces \cite{Maldacena:2000mw}. As expected, these form a subclass of the geometries \eqref{eq:LLM-metric}, but they are also vacua of the 7-dimensional maximally supersymmetric SO(5) gauged supergravity obtained by truncating 11-dimensional supergravity on a $S^4$ \cite{Nastase:1999cb}. In this section, we describe this class of vacua and analyse consistent truncations around them in terms of the generalised parallelisation of $S^4$ used to construct SO(5) gauged supergravity as a generalised Scherk-Schwarz reduction. A particular example of such a consistent truncation was analysed in \cite{Cassani:2019vcl} using generalised geometry and constructed using different methods in \cite{Cheung:2019pge}. Here we will show that this truncation is the most general belonging to this class.

\subsection{$S^4$ reduction and embedding into $E_{6(6)}$}
As shown in \cite{Lee:2014mla,Hohm:2014qga}, the $S^4$ is generalised parallelisable in $\SL{5}$ ExFT. This means that its generalised tangent bundle admits a globally defined frame, which can be used to define a consistent truncation preserving all supersymmetries. This generalised parallelisation is given by the following objects in the \textbf{10}
\begin{equation}
E_{IJ\,(\textbf{10})}=R^{-1}\,v_{IJ}+R^2\,\sigma_{IJ}+R^{-1}\,\iota_{v_{IJ}} A_{(3)}\,,
\end{equation}
where $I,J=1\dots 5$, $R$ is the radius of the 4-sphere, and the objects $v_{IJ}$ and $\sigma_{IJ}$ are defined in appendix \ref{sec:app-spheres}. The three form $A_{(3)}$ is defined such that
\begin{equation}\label{eq:A3-SL(5)}
	dA_{(3)} = 3\, R^3\, vol_{S^4}\,.
\end{equation}
Furthermore, one can also define global frames for other generalised bundles of the sphere in other representations of SL(5). For instance, the paralellisations of the bundles transforming in the $\bar{\textbf{5}}$ and the \textbf{5} are defined by
\begin{equation}
\begin{split}
E_{I,(\bar{\textbf{5}})}&=R\,d\mathbb{Y}_I-R^4\,Y_I\,vol_{S^4} +R\, d\mathbb{Y}_I\wedge A_{(3)}\,,\\
E_{I,(\textbf{5})}&=-R^3\,\beta_I + \mathbb{Y}_I\, A_{(3)}+\mathbb{Y}_I\,,
\end{split}
\end{equation}
where we refer again to appendix \ref{sec:app-spheres} for the definitions of the objects appearing. These parallelisations satisfy
\begin{equation}
E_I{}^{[\mbf{a}}E_J{}^{\mbf{b}]}=\frac{1}{2}\,|R^2\,g_{S^4}|^{1/2}\,E_{IJ}{}^{\mbf{ab}}\,,
\end{equation}
where $\mbf{a}, \mbf{b} = 1, \ldots, 5$ are fundamental $\SL{5}$ indices. The generalised paralellisation generates a SO(5) algebra, namely
\begin{equation}
	\gL_{E_{IJ\,(\mathbf{10})}}E_{KL\,(\mathbf{10})} = -\frac{1}{R}\,\left(\delta_{IK}E_{JL\,(\mathbf{10})} - \delta_{IL}E_{JK\,(\mathbf{10})} - \delta_{JK}E_{IL\,(\mathbf{10})} + \delta_{JL}E_{IK\,(\mathbf{10})}\right) \,.
\end{equation}
As discussed in appendix \ref{sec:app-spheres}, this symmetry can be broken into SU(2)$\times$U(1) by choosing the embedding coordinates for the $S^4$ as \eqref{eq:S4toS2S1}. Splitting the index $I=(A,i)$, we define the following generalised objects
\begin{equation}\label{eq:SU(2)xU(1)-from-SO(5)}
\begin{split}
E_{A,(\mathbf{10})}&=\frac{1}{2}\epsilon_{ABC}E_{BC,(\mathbf{10})}\,,\qquad E_{A,(\mathbf{5})}=E_{I=A,(\mathbf{5})}
\,,\qquad E_{A,(\bar{\mathbf{5}})}=E_{I=A,(\bar{\mathbf{5}})}\,,\\
E_{S^1,(\mathbf{10})}&=\frac{1}{2}\epsilon_{ij}E_{ij,(\mathbf{10})}\,,\qquad E_{i,(\mathbf{5})}=E_{I=i,(\mathbf{5})}
\,,\qquad E_{i,(\bar{\mathbf{5}})}=E_{I=i,(\bar{\mathbf{5}})}\,,
\end{split}
\end{equation}
where now the objects $E_{A,(\mathbf{10})}$ and $E_{S^1,(\mathbf{10})}$ generate the SU(2) and the U(1) algebras. The rest transform naturally under their action, namely
\begin{equation}
\begin{split}
\mathcal{L}_{E_{A,(\mathbf{10})}}E_{B,(\mathbf{10})}&=-\frac{1}{R}\,\epsilon_{ABC}E_{C,(\mathbf{10})}\,,\qquad
\mathcal{L}_{E_{A,(\mathbf{10})}}E_{B,(\mathbf{5})}=-\frac{1}{R}\,\epsilon_{ABC}E_{C,(\mathbf{5})}\,,\\
\mathcal{L}_{E_{A,(\mathbf{10})}}E_{B,(\bar{\mathbf{5}})}&=-\frac{1}{R}\,\epsilon_{ABC}E_{C,(\bar{\mathbf{5}})}\,,\\
\mathcal{L}_{E_{S^1,(\mathbf{10})}}E_{i,(\mathbf{5})}&=-\frac{1}{R}\,\epsilon_{ij}E_{j,(\mathbf{5})}\,,\qquad
\mathcal{L}_{E_{S^1,(\mathbf{10})}}E_{i,(\bar{\mathbf{5}})}=-\frac{1}{R}\,\epsilon_{ij}E_{j,(\bar{\mathbf{5}})}\,.
\end{split}
\end{equation}
In terms of the objects on the 2- and 1- spheres, the gauge field in \eqref{eq:A3-SL(5)} can be written as
\begin{equation}
A=r^3\, vol_{S^2}\wedge vol_{S^1}\,.
\end{equation}

\subsection{Half maximal generalised structures}
Given the parallelisation of the $S^4$ described above and the decomposition of the $E_{6(6)}$ covariant objects into SL(5) $\times$ GL(2) objects described in appendix \ref{sec:app-SL5}, we can now construct the most general geometries with 16 supercharges of the local form AdS$_5\times \Sigma\times S^4$, with $\Sigma$ a Riemann surface. The most general Ansatz compatible with the symmetries is
\begin{equation}
\begin{split}
J_A&=5^{1/6}\sqrt{2}\,E_{A,(\mathbf{10})}+f\,E_{A,(\mathbf{5})}\otimes vol_{\Sigma}+E_{A,(\bar{\mathbf{5}})}\otimes \varphi\,,\\
J_i&= h\, E_{i,(\mathbf{5})}\otimes vol_{\Sigma}+\tilde h\, \epsilon E_{i,(\mathbf{5})}\otimes vol_{\Sigma}+E_{i,(\bar{\mathbf{5}})}\otimes \xi+\epsilon E_{i,(\bar{\mathbf{5}})}\otimes \tilde\xi\,,\\
\hat{K}&=5^{1/6}\sqrt{2}\,E_{S^1,(\mathbf{10})}\,,
\end{split}
\end{equation}
where $f$, $h$ and $\tilde{h}$ are functions of $\Sigma$ and $\varphi$, $\xi$ and $\tilde\xi$ are 1-forms on $\Sigma$. Imposing the algebraic constraints \eqref{eq:SO5-structure-cond} they become
\begin{equation}\label{eq:gen-structures-S4}
\begin{split}
J_A&=5^{1/6}\sqrt{2}\, E_{A,(\mathbf{10})}+R^2\,E_{A,(\mathbf{5})}\otimes (\xi\wedge\tilde\xi)+R\,E_{A,(\bar{\mathbf{5}})}\otimes \varphi\,,\\
J_i&= 5^{1/6}\sqrt{2}\,R^2\,\Bigl[E_{i,(\mathbf{5})}\otimes(\tilde\xi\wedge\varphi)+\epsilon E_{i,(\mathbf{5})}\otimes (\xi\wedge\varphi)\Bigr] - 5^{1/6}\sqrt{2}\,R\,\Bigl[E_{i,(\bar{\mathbf{5}})}\otimes \xi-\epsilon E_{i,(\bar{\mathbf{5}})}\otimes \tilde\xi\Bigl]\,,\\
\hat{K}&= 5^{1/6}\sqrt{2}\, E_{S^1,(\mathbf{10})}\,,
\end{split}
\end{equation}
where we have rescaled the forms  $\varphi$, $\xi$ and $\tilde\xi$ for later convenience, and the condition $K\wedge\hK>0$ implies
\begin{equation}
	\xi\wedge\tilde\xi>0\,.
\end{equation}
Finally, the differential constraints \eqref{eq:Gen-weakly-integrability-cond} impose
\begin{equation}\label{eq:conditions-SL(5)-section}
d\varphi=-\xi\wedge\tilde\xi \,,\qquad d\xi=-\varphi\wedge \tilde\xi\,,\qquad d\tilde\xi=\varphi\wedge \xi\,.
\end{equation}
and the result can be embedded into the general one \eqref{eq:structures-final} obtained in section \ref{sec:AdS5-vacua} by taking
\begin{equation}\label{eq:dictionary-Mtheory-SL5}
\begin{split}
p&=5^{1/6}\sqrt{2}\,R^2\,r\,,\qquad \qquad \qquad \,\chi=-5^{1/6}\sqrt{2}\,R^2\,r\,\varphi\,,\\ \nu_1&=5^{1/6}\sqrt{2}\,R^2\,\sqrt{1-r^2}\,\xi\,,\qquad \nu_2=5^{1/6}\sqrt{2}\,R^2\,\sqrt{1-r^2}\,\tilde\xi\,.
\end{split}
\end{equation}

\subsection{AdS$_5$ vacua fields}
As we did in the general case, we can obtain the geometry whose $G$-structures are given by \eqref{eq:gen-structures-S4} by constructing the generalised metric \eqref{eq:Gen-Metric-from-structures}  and using the dictionary \eqref{eq:dictionary-E6}. We obtain the following configuration
\begin{equation}\label{eq:MN-solution}
\begin{split}
ds^2&=\frac{1}{2}\,W^{1/3}\Bigl(4\,ds^2_{AdS_5}+\\
&\qquad 2\,R^2\,(\xi\otimes\xi+\tilde\xi\otimes\tilde\xi)+\frac{2\,R^2}{1-r^2}\,dr\otimes dr+\frac{2\,R^2}{W}\,r\,ds^2_{S^2}+\frac{4\,R^2}{W}(1-r^2)(d\beta+\varphi)\otimes  (d\beta+\varphi)\Bigr)\,,\\
C_{(3)}&=\frac{4\,R^3\,r^3}{W}(d\beta+\varphi)\wedge vol_{S^2}\,,
\end{split}
\end{equation}
with
\begin{equation}
W=1+r^2\,.
\end{equation}
Next, as we did for the general case, we can use diffeomorphisms on the Riemann surface to fix a gauge for the forms $\xi$ and $\tilde\xi$. Analogous to what we discussed before, we can take a gauge where
\begin{equation}
\begin{split}
\xi&=e^{\sigma/2}(\cos\Theta dx^1+\sin\Theta dx^2)\,,\\
\tilde\xi&=e^{\sigma/2}(-\sin\Theta dx^1+\cos\Theta dx^2)\,,
\end{split}
\end{equation}
where $\Theta$ and $\sigma$ are two functions on the Riemann surface. With this choice, the last two conditions in \eqref{eq:conditions-SL(5)-section} are solved by taking
\begin{equation}
\varphi=\frac{1}{2}(\epsilon_{ij}\partial_j\sigma)dx^i-d\Theta\,,
\end{equation}
and, as in the general case, we observe that $\Theta$ can be completely removed from the geometry by a local redefinition of the U(1) coordinate. Finally, the first condition in \eqref{eq:conditions-SL(5)-section} imposes that $\sigma$ must satisfy the equation
\begin{equation}\label{eq:Toda-eq-for-sigma}
\partial_1^2\sigma+\partial_2^2\sigma-2\,e^\sigma=0\,,
\end{equation}
which is related to the Toda equation \eqref{eq:Toda-equation} of the general case by taking
\begin{equation}\label{eq:MN-in-LLM}
e^{D(x_1,x_2,r)}=(1-r^2)\,e^{\sigma(x_1,x_2)}
\end{equation}
the coordinate $r$ playing the role of the coordinate $y$ in section \ref{sec:AdS5-vacua}. A particular solution to equation \eqref{eq:Toda-eq-for-sigma} is given by
\begin{equation}
	e^{\sigma(x_1,x_2)}=\frac{4}{(1-(x_1^2+x_2^2))^2}\,,
\end{equation} 
whereby the geometry \eqref{eq:MN-solution} becomes the Maldacena-Nu\~{n}ez configuration \cite{Maldacena:2000mw}.

\subsection{Consistent truncations with vector multiplets}
In section \ref{sec:vector-multiplets}, we argued that geometries described by a function $D$ of the form \eqref{eq:MN-in-LLM} allowed for a consistent truncation with three vector multiplets around it, with gauge group ISO(3) $\times \UO$. This truncation can also be embedded into maximal 7-dimensional SO(5) gauged supergravity, as explicitly constructed in \cite{Cheung:2019pge}, and also analysed in \cite{Cassani:2019vcl}. In this section, we will argue that this truncation and the minimal one are the only half-maximal consistent truncations one can construct around half-maximal AdS$_5$ vacua of 7-dimensional SO(5) gauged supergravity. We stress, however, that this does not necessarily exclude truncations around these vacua that cannot be embedded into the 7-dimensional theory. To analyse the existence of such truncations, we refer to the general discussion of section \ref{sec:vector-multiplets}, where one should take $D$ of the form \eqref{eq:MN-in-LLM}.

As reviewed in \ref{sec:vector-multiplets-review}, in order to have consistent truncations the internal geometry should admit extra generalised vectors forming a generalised $\Spin{5-n} \subset \USp{4}$ structure. Since we want the consistent truncation to come from 7-dimensional gauged supergravity, the generalised $\Spin{5-n} \subset \USp{4}$ structure must not deform the generalised parallelisation of the $S^4$. Therefore, the $\SL{5}$ part of the generalised structures must be written in terms of \eqref{eq:SU(2)xU(1)-from-SO(5)}.

We start by analysing the case of a single vector multiplet, transforming as a singlet of the R-symmetry group. From \eqref{eq:SU(2)xU(1)-from-SO(5)}, we observe that the only object transforming as a singlet is $E_{S^1, (\mathbf{10})}$. However, the generalised wedge product of this object with itself is zero, so it cannot be used to construct a single generalised vector satisfying the conditions \eqref{eq:vec-mult-algebraic-cond-rev}. Given that this is the only possibility, we conclude that truncations with a single vector multiplet from 7-dimensional SO(5) gauged supergravity do not exist. 

Finally, we analyse the case of two extra vector multiplets transforming as a complex U(1) representation of charge $q$. As in section \ref{sec:vector-multiplets}, we can construct them by writing the most general ansatz compatible with the symmetries. In this case, the most general ansatz compatible with the algebraic conditions \eqref{eq:vec-mult-algebraic-cond-rev} is
\begin{equation}
\begin{split}
\bar J_{\bar{\imath}}&= 5^{1/6}\sqrt{2}\,R^2\,\Bigl[E_{i,(\mathbf{5}),(q)} \otimes(\tilde\chi\wedge\varphi) +\epsilon E_{i,(\mathbf{5}),(q)}\otimes (\chi\wedge\varphi)\Bigr] \\
&\quad- 5^{1/6}\sqrt{2}\,R\,\Bigl[E_{i,(\bar{\mathbf{5}}),(q)}\otimes \chi-\epsilon E_{i,(\bar{\mathbf{5}}),(q)}\otimes \tilde\chi\Bigl]\,,
\end{split}
\end{equation}
where $E_{i,(..),(q)}$ are U(1) representations of charge $q$ constructed analogously to the forms $\omega_{(q)}{}^i$ in appendix \ref{sec:app-spheres}. The forms $\chi$ and $\tilde\chi$ satisfy
\begin{equation}
	\tilde\chi\wedge\xi = \chi\wedge\tilde\xi\,,\qquad  \chi\wedge\xi=-\tilde\chi\wedge\tilde\xi\,,\qquad \chi\wedge\tilde\chi=-\xi\wedge\tilde\xi\,,
\end{equation}
which can be solved in general by taking 
\begin{equation}
	\chi=V_1\xi+V_2\tilde\xi\,,\qquad \tilde\chi = \pm(V_2\xi - V_1\tilde\xi)\,,
\end{equation}
with $V_1$ and $V_2$ two functions of the Riemann surface satisfying $V_1^2+V_2^2=1$. Given this solution, and unlike in the general case discussed in section \ref{s:Doublet}, the differential conditions \eqref{eq:vec-mult-diff-cond-rev} can only be solved if the charge is $q=1$. In this situation the forms $\chi$ and $\tilde\chi$ need to satisfy
\begin{equation}
	d\tilde\chi =-\chi\wedge\varphi\,,\qquad d\chi=\tilde\chi\wedge\varphi\,,
\end{equation}
and one can show, in a very similar way to the discussion of section \ref{s:TripletDoublet}, that these equations cannot be solved given that the function $\sigma$ characterising the vacuum satisfies equation \eqref{eq:Toda-eq-for-sigma}. This concludes our proof that the only possible consistent truncations around half-maximal AdS$_5$ vacua in 7-dimensional SO(5) gauged supergravity are the minimal truncation keeping only the gravitational supermultiplet or the consistent truncation with three vector multiplets and ISO$(3) \times \UO$ gauging constructed in \cite{Cheung:2019pge,Cassani:2019vcl}.

\section{Conclusions}
In this paper, we studied half-maximal AdS$_5$ vacua of M-theory and their consistent truncations. We showed that the half-maximal AdS$_5$ vacua are captured in ExFT by generalised $\USp{4}$ structures whereby the Toda equation arises from the ``weak integrability condition'' that is equivalent to the BPS condition. Moreover, we showed that ExFT allows us to fully classify all possible consistent truncations to half-maximal gauged supergravity with arbitrary number of vector multiplets. In particular, all half-maximal AdS$_5$ vacua admit a consistent truncation keeping only the gravitational supermultiplet \cite{Gauntlett:2007sm}, and these arise immediately from the ExFT formulation of the AdS vacua. Moreover, when the half-maximal AdS$_5$ vacuum admits a generalised $\Spin{5-n} \subset \USp{4}$ structure satisfying certain differential conditions, then we can define a consistent truncation with $n$ vector multiplets.

By working systematically through all possible cases, we showed that the largest possible consistent truncation of the half-maximal AdS$_5$ vacua contain 3 vector multiplets, with the possible gaugings ISO$(3) \times$ U(1) or $\SU{2} \times \UO$, where in the latter the $\UO$ is embedded either as a subgroup of $\SO{5} \subset \SO{5,3}$ or as a linear combination of $\UO \subset \SO{5} \subset \SO{5,3}$ and $\UO \subset \SO{3} \subset \SO{5,3}$. We were able to show that the ISO$(3) \times \UO$ gauging arises if and only if the internal space is a $S^4$ warped over a Riemann surface. This consistent truncation has previously been constructed in \cite{Cheung:2019pge} and studied from a generalised geometry perspective in \cite{Cassani:2019vcl}.

We also derived differential conditions which must be satisfied by the AdS$_5$ vacua in order to admit a consistent truncation with one or two vector multiplets. The resulting gaugings are $\SU{2} \times \UO$ with the $\UO$ again possibly embedded in two different ways for the case of two vector multiplets. Any other 5-dimensional half-maximal gauged supergravity with AdS$_5$ vacua preserving 16 supercharges cannot be uplifted to M-theory. Therefore, our results can be seen as a higher-dimensional completion of the 5-dimensional analysis of the possible gaugings giving rise to half-maximal AdS$_5$ vacua in \cite{Louis:2015dca}.

In section \ref{sec:AdS-SL(5)}, we analysed in more detail truncations around vacua whose internal space is locally a $S^4$ fibration over a Riemann surface. As mentioned above and previously shown in \cite{Cheung:2019pge,Cassani:2019vcl}, these admit an ISO$(3)\times\UO$ truncation around them, which can be constructed via the SO$(5)$ gauged supergravity obtained by reducing 11-dimensional supergravity on $S^4$. We argue that this is the only consistent truncation with vector multiplets that can be obtained in this way.

A natural follow-up question to our work is to better understand the differential equations that arise in the classification of consistent truncations with vector multiplets and to find examples of AdS$_5$ vacua satisfying these conditions, or alternatively, proving that no such vacua can exist. This would finally settle the question of which 5-dimensional half-maximal gauged supergravities with half-maximal AdS$_5$ vacua can be uplifted to 11-dimensional supergravity. A particularly interesting example would be the half-maximal theory analysed in \cite{Bobev:2018sgr} which contains RG flows between half-maximal and ${\cal N}=2$ AdS$_5$ vacua. This theory belongs to the class we studied in section \ref{s:Doublet}. Therefore, if the conditions \eqref{eq:alg-cond-Doublet-VM-1}, \eqref{eq:diff-cond-Doublet-VM-1} and \eqref{eq:diff-cond-Doublet-VM-2} can be solved, the flow constructed in \cite{Bobev:2018sgr} can be immediately uplifted using our results.

Finally, ExFT has recently been developed as a powerful tool to study the full Kaluza-Klein spectrum around vacua of gauged supergravities \cite{Malek:2019eaz,Malek:2020yue}. In these methods, consistent truncations play an important role, since they allow us to study all vacua lying in the same truncation, even those with less or no (super-)symmetry remaining. So far, the method has been restricted to consistent truncations preserving all supercharges \cite{Malek:2020mlk,Guarino:2020flh,Bobev:2020lsk}. However, a systematic constructions of half-maximal truncations can open the door to adapt these methods to consistent truncations preserving fewer supersymmetries. This might give access to the Kaluza-Klein spectra around vacua of half-maximal gauged supergravities.

\section*{Acknowledgements}
The authors thank Nikolay Bobev, Davide Cassani, Carlos Nu\~{n}ez, Michela Petrini, Hagen Triendl and Daniel Waldram for useful discussions. EM is supported by the Deutsche Forschungsgemeinschaft (DFG, German Research Foundation) via the Emmy Noether program ``Exploring the landscape of string theory flux vacua using exceptional field theory'' (project number 426510644). The work of VVC is supported by the Alexander von Humboldt Foundation via a Feodor Lynen fellowship. VVC also thanks the Albert-Einstein-Institute, Potsdam for support while part of this project was completed.

\appendix

\section{$E_{6(6)}$ ExFT and 11-dimensional supergravity}\label{sec:app-11d-section}
In this appendix we collect some facts about $E_{6(6)}$ ExFT and how 11-dimensional supergravity is embedded into it. By taking the 11-dimensional solution to the section constraint, the $E_{6(6)}$ symmetry is broken to $\GL{6}$. Under such a breaking, a generalised vectors $J\in\textbf{27}$ with weight $1/3$ and a generalised vector $K\in\overline{\textbf{27}}$ with weight $2/3$ decompose as
\begin{equation}\label{eq:splitting-27-11d}
\begin{split}
J &= v + \lambda_{(2)} + \lambda_{(5)} \,, \\
K &= \omega_{(1,6)} + \omega_{(3)} + \omega_{(1)}\,,
\end{split}
\end{equation}
where $v$ is a vector and $\lambda_{(n)}$ and  $\omega_{(n)}$ are forms of degree $n$.  The object $\omega_{(1,6)}$ transforms as a section of $T^*M \otimes \Lambda^6 T^*M$. In components we write
\begin{equation}
\begin{split}
J^N&=(J^n, J_{n_1n_2}, J_{n_1\dots n_5})\,,\\
K_N&=(K_{\bar{n}}, K_{n_1n_2n_3}, K_{n})\,.
\end{split}
\end{equation}
The group $E_{6(6)}$ has a totally symmetric invariant tensor $d_{MNP}$, whose non-zero components  are
\begin{equation}\label{eq:dMNP}
\begin{split}
d^m{}_{np,qrstv} &= \sqrt{\frac{2}{5}}\, \delta^m_{[n}\, \epsilon_{p]qrstv} \,, \\
d_{mn,pq,rs} &= \frac{1}{\sqrt{10}}\, \epsilon_{mnpqrs} \,,
\end{split}
\end{equation}
where we use the conventions where the tensor $d^{MNP}$ has the same numerical factors as $d_{MNP}$, and it is normalised as\footnote{\label{ftnote:conventions}Note that the numerical factors of $d_{MNP}$ given in \eqref{eq:dMNP} differ from those in \cite{Hohm:2013vpa}. The reason is a different summation convention: in ours, we add a factorial factor every time we sum over anti-symmetrised indices, namely,
	\begin{equation*}
	V_1{}^NV_{2\,N}=V_1{}^nV_{2\,n}+\frac{1}{2!}V_1{}^{n_1n_2}V_{2\,n_1n_2}+\frac{1}{5!}V_1{}^{n_1\dots n_5}V_{2\,n_1\dots n_5}\,,
	\end{equation*} 
	whereas their summation convention is without the factorial factors.}
\begin{equation}
d_{NPQ}\,d^{MPQ} = \delta_N^M\,.
\end{equation}
Furthermore, it satisfies  the cubic identity
\begin{equation}
d_{S(MN}d_{PQ)T}d^{STR}=\frac{2}{15}\delta_{(M}^R d_{N)PQ}\,.
\end{equation}
With the splitting \eqref{eq:splitting-27-11d}, the wedge products \eqref{eq:wedge27-def} and \eqref{eq:wedge27bar-def} decompose as
\begin{equation}
\begin{split}
\left( \J_u \wedge \J_v \right)_{(1,6)} &= \sqrt{\frac{2}{5}}\, dx^m \otimes \lambda_{(u|\,m,(1)}\wedge \lambda_{|v),(5)} \,, \\
\left( \J_u \wedge \J_v \right)_{(4)} &=\sqrt{\frac{2}{5}} \left(\frac{1}{2} \lambda_{u\,(2)} \wedge \lambda_{v\,(2)}-\iota_{v_{(u}} \lambda_{v)\,(5)}   \right) \,, \\
\left( \J_u \wedge \J_v \right)_{(1)} &=-\sqrt{\frac{2}{5}}\, \iota_{v_{(u}} \lambda_{v)\,(2)} \,, \\
\K \wedge \hK &= (\iota_{v} \omega_{(1,6)})_{(6)} + \lambda_{(2)} \wedge \omega_{(4)} + \lambda_{(5)}\wedge \omega_{(1)} \,,
\end{split}
\end{equation}
where $J_u, \, J_v,\, \hat{K} \in\textbf{27}$ and $K\in\overline{\textbf{27}}$. The object $\lambda_{u\,m,(1)}$ in the first line is defined as
\begin{equation}
\lambda_{m,(1)}=\lambda_{mp}\,dx^p\,,
\end{equation}
where $\lambda_{mp}$ are the components of $\lambda_{(2)}$, i.e. $\lambda_{(2)}=\frac{1}{2!}\lambda_{mp}dx^m\wedge dx^p$. Also, the object $ (\iota_{v} \omega_{(1,6)})_{(6)}$ is a contraction of the vector $v$ with the 1-form part of $\omega_{(1,6)}$, thus obtaining a 6-form:
\begin{equation}
(\imath_{V} \omega_{(1,6)})_{(6)}=V^m \omega_{m,(6)}\,,
\end{equation}
The generalised Lie derivative \eqref{eq:GenLieDerivative} with parameter $J_u\in\textbf{27}$ acting on a vector $J_v\in\textbf{27}$ decomposes under the split \eqref{eq:splitting-27-11d} as
\begin{equation}
\begin{split}
\left(\mathcal{L}_{J_u}J_v\right)_{(v)}&=L_{v_u}v_v\,,\\
\left(\mathcal{L}_{J_u}J_v\right)_{(2)}&=L_{V_u}\lambda_{v,(2)}-\iota_{V_v}d\lambda_{u,(2)}\,,\\
\left(\mathcal{L}_{J_u}J_v\right)_{(5)}&=L_{V_u}\rho_{v,(5)}-\iota_{V_v}d\rho_{u,(5)}-\lambda_{v,(2)}\wedge d\lambda_{u,(2)}\,.
\end{split}
\end{equation}

\subsubsection*{11-dimensional fields from the ExFT generalised metric}
As mentioned above, all purely internal degrees of freedom of the 11-dimensional supergravity are encoded into the generalised metric $\mathcal{M}_{MN}$.  For instance, the components $\gM_{\bar{m}\bar{n}}$ and $\gM_{\bar m}{}^{kl}$ are \cite{Hohm:2013vpa}\footnote{Note that the numerical factors are different from those appearing in \cite{Hohm:2013vpa}. The reason is the different summation convention as explained in footnote \ref{ftnote:conventions}.}
\begin{equation}\label{eq:dictionary-E6}
\begin{split}
\gM_{\bar{m}\bar{n}}&=(\det g)^{-2/3}g_{mn}\,,\\
\gM_{\bar m}{}^{kl}&=\frac{1}{2}(\det g)^{-2/3}g_{mn}\epsilon^{nklpqr}\,C_{pqr} \,,
\end{split}
\end{equation}
where the barred indices in the left hand side indicate that they correspond to the (dualised) 5-form components.

\section{Spheres conventions}\label{sec:app-spheres}
We summarise here the notation and conventions for the spheres of different dimensions appearing in the paper. In general, one defines a unit radius sphere $S^d\subset\mathbb{R}^{d+1}$ in terms of $d+1$ functions $\mathcal{Y}_{\mathcal{I}}$, $\mathcal{I}=1,\ldots, d+1$ satisfying 
\begin{equation}\label{eq:Y2-Sd}
	\mathcal{Y}_{\mathcal{I}}\, \mathcal{Y}^{\mathcal{I}}=1\,.
\end{equation}
The round metric on the sphere and its volume form are
\begin{equation}
	ds^2=d\mathcal{Y}_{\mathcal{I}}\otimes d\mathcal{Y}^{\mathcal{I}}\,,\qquad
	vol_{S^d}=\frac{1}{d!}\epsilon_{\mathcal{I}_1\dots\mathcal{I}_{d+1}} \mathcal{Y}^{\mathcal{I}_1} d\mathcal{Y}^{\mathcal{I}_2}\wedge\dots\wedge d\mathcal{Y}^{\mathcal{I}_{d+1}}\,.
\end{equation}
This metric space has $d(d+1)/2$ Killing vectors which can be labelled by an antisymmetric index $[\mathcal{I}\mathcal{J}]$ and,  given a local coordinate base $x^{\mathbf{i}}$, with $\mathbf{i}=1\dots d$, can be written as
\begin{equation}\label{eq:Killing-vecs-Sd}
	V_{[\mathcal{I}\mathcal{J}]}{}^{\mathbf{i}}
	=2\,g^{\mathbf{i}\mathbf{j}}\mathcal{Y}_{[\mathcal{I}}\partial_{\mathbf{j}}\mathcal{Y}_{\mathcal{J}]}\,,
\end{equation}
where $g^{\mathbf{i}\mathbf{j}}$ is the inverse metric on the sphere.

\subsubsection*{The 2-sphere}
The 2-sphere $S^2$ is defined by three functions $Y^A$ with $A=1,2,3$ and satisfying relation \eqref{eq:Y2-Sd}. The three Killing vectors \eqref{eq:Killing-vecs-Sd} can be organised into
\begin{equation}
	v_A=\frac{1}{2}\epsilon_A{}^{BC}v_{BC}\,.
\end{equation}
Furthermore, we also introduce the 1-forms which are Hodge-dual to $dY_A$, given by
\begin{equation}
\theta_A=\epsilon_{ABC}Y^BdY^C\,.
\end{equation}
Together with the other vector and forms on the 2-sphere these satisfy the following relations
\begin{equation}
	\iota_{v_A}\theta_B=\delta_{AB}-y_Ay_B\,,\qquad Y_AdY^A=Y_A\theta^A=y_Av^A=0\,.
\end{equation}
Finally, all objects transform in a natural way under the SO(3) action generated by the Killing vectors, namely
\begin{equation}
	L_{v_A}v_B=-\epsilon_{ABC}v^C\,,\quad 	L_{v_A}y_B=-\epsilon_{ABC}y^C\,,\quad 	L_{v_A}dY_B=-\epsilon_{ABC}dY^C\,,\quad L_{v_A}\theta_B=-\epsilon_{ABC}\theta^C\,.
\end{equation}
\subsubsection*{The 1-sphere}
The circle $S^1$ can be described in terms of two functions $w^i$, $i=1,2$ satisfying \eqref{eq:Y2-Sd}. There is a single Killing vector,
\begin{equation}
	v_{S^1}=\frac{1}{2}\epsilon^{ij}v_{ij}\,.
\end{equation}
The functions $w^i$ and their external derivatives transform naturally under its action, namely
\begin{equation}
L_{v_{S^1}}w^i=-\,\epsilon^{ij}\,w_{j}\,,\qquad
L_{v_{S^1}}dw^i=-\,\epsilon^{ij}\,dw_{j}\,,
\end{equation}
Given the set of functions $w^i$, one can construct another set of functions
\begin{equation}
	\epsilon w^i=\epsilon^{ij} w_j\,,
\end{equation}
which under the action of $v_{S^1}$ transform in the same way as $w^i$. Some useful relations are
\begin{equation}
	dw^i=-\epsilon w^i vol_{S^1}\,,\qquad d(\epsilon w^i)=w^i vol_{S^1}\,.
\end{equation}
Finally, one can also construct functions transforming as U(1) representation of arbitrary integer weight under the action of $v_{S^1}$. For instance, if one takes the defining functions of the circle to be 
\begin{equation}
w_i=(\cos\beta,\sin\beta)\,,
\end{equation}
where $\beta\in [0,2\pi)$ is the angle on the circle, one can construct the functions
\begin{equation}
w_{(q)}{}^i=(\sin(q\,\beta),\,\cos(q\,\beta)\,)\,,
\end{equation}
with $q\in\mathbb{Z}$, which transform under the action of  $v_{S^1}$ as
\begin{equation}
L_{v_{S^1}}w_{(q)}{}^i=-q\,\epsilon^{ij}\,w_{(q)\,j}\,,\qquad
L_{v_{S^1}}dw_{(q)}{}^i=-q\,\epsilon^{ij}\,dw_{(q)\,j}\,,
\end{equation}
and analogous with $\epsilon w_{(q)}{}^i$.

\subsubsection*{The 4-sphere}
We describe the four-sphere $S^4$ in terms of five functions $\mathbb{Y}^I$ satisfying \eqref{eq:Y2-Sd}. Other forms that are relevant for our discussion are
\begin{equation}
\begin{split}
\sigma_{IJ}&=\frac{1}{2}\epsilon_{I J K_1K_2K_3}\mathbb{Y}^{K_1} d\mathbb{Y}^{K_2}\wedge d\mathbb{Y}^{K_3}\,,\\
\beta_{I}&=\frac{1}{3!}\epsilon_{IJ_1\dots J_4}\mathbb{Y}^{J_1} d\mathbb{Y}^{J_2}\wedge d\mathbb{Y}^{J_3}\wedge d\mathbb{Y}^{J_4}\,.
\end{split}
\end{equation}
The 4-sphere can be described in terms of the defining functions 2- and 1- spheres by taking
\begin{equation}\label{eq:S4toS2S1}
	\mathbb{Y}^I = (-r\,Y^A, \sqrt{1-r^2}\, w^i)\,,
\end{equation}
where $Y^A$ and $w^i$ are the coordinates for the 2- and the 1- spheres defined above, namely $Y_AY^A=w_iw^i=1$, and $r\in [0,1)$ is an extra coordinate. The minus sign in front of the $S^2$ part ensures that the 4-, 2- and 1- spheres have all positive-definite volume forms.

\section{$\SL{5}$ ExFT and its embedding in $E_{6(6)}$}\label{sec:app-SL5}
The relevant group for constructing an exceptional field theory adapted to a 7+4 splitting of 11-dimensional supergravity is SL(5). In this ExFT \cite{Berman:2010is}, the generalised internal coordinates transform in the \textbf{10} of SL(5). Other relevant representations appearing in its construction are the \textbf{5} and the $\overline{\textbf{5}}$. 

By taking the 11-dimensional supergravity solution of the section constraint, the SL(5) group is broken to GL(4). The objects transforming in the \textbf{10}, \textbf{5} and $\overline{\textbf{5}}$ with weights $\frac15$, $\frac35$ and $\frac25$, respectively, under the generalised Lie derivative decompose as
\begin{equation}
	\begin{split}
	V_{\textbf{10}} &= v+\lambda_{(2)}\,,\\
	V_{\textbf{5}} &= \lambda_{(3)} + \lambda_{(0)}\,,\\
	V_{\overline{\textbf{5}}} &= \lambda_{(1)} + \lambda_{(4)}\,,
	\end{split}
\end{equation}
 where $v$ is a vector and $\lambda_{(n)}$ an $n$-form. 
 
\subsubsection*{Embedding into $E_{6(6)}$}
In appendix \ref{sec:app-11d-section} we discussed that taking the 11-dimensional supergravity solution to the section constraint breaks the group $E_{6(6)}$ to GL(6). Here we solve the section conditions in a different way, adapted to the case where the internal space has itself a fibration structure with a 4+2 split. In particular, this structure breaks $E_{6(6)}$ to SL(5) $\times$ GL(2). Then, the usual 11-dimensional solution to the section constraint is recovered by taking the SL(5) solution to the section constraint and letting fields depend also on the GL(2) coordinates. However, for certain calculations it is useful to keep the SL(5) covariance, as we will use now. Under the splitting $E_{6(6)}\longrightarrow \SL{5} \times \GL{2}$,  an element $V^N$ in the \textbf{27} representation of $E_{6(6)}$ decomposes as
\begin{equation}
V=V_{\textbf{10},(0)}+V_{\textbf{5},(2)}+V_{\bar{\textbf{5}},(1)}+V_{\textbf{0},(v)}\,,
\end{equation} 
where the bold subscripts indicate the SL(5) representation and the subscript $(n)$ that the object transforms like a GL(2) $n$-form (or vector for $(v)$). In components,
\begin{equation}
V^N=(V^{\mbf{ab}}, V^{\mbf{a}}{}_{\alpha\beta}, V_{\mbf{a}\,\alpha}, V^\alpha)\,,
\end{equation} 
where $\mbf{a}, \mbf{b}, = 1, \ldots, 5$ label the fundamental $\SL{5}$ indices and $\alpha, \beta = 1, 2$ label the fundamental $\GL{2}$ indices. The $E_{6(6)}$ invariant $d_{MNP}$ is given by 
\begin{equation}
\begin{split}
d_{\mbf{ab}\,\mbf{cd}\,\mbf{e}}{}^{\alpha\beta}&=\frac{1}{\sqrt{10}}\epsilon_{\mbf{abcde}}\left(\frac{1}{2}\epsilon^{\alpha\beta}\right)\,,\qquad d_{\mbf{ab}}{}^{\mbf{c}\, 1\,\mbf{d}\,2}=-\sqrt{\frac{2}{5}}\,\delta_{\mbf{ab}}^{\mbf{cd}} \,, \\
d_{\mbf{a}}{}^{\alpha\beta\,\mbf{b}}{}_\gamma{}^\sigma&=-\frac{1}{\sqrt{10}}\delta_{\mbf{a}}^{\mbf{b}}\,\delta_\gamma^\sigma\left(\frac{1}{2}\epsilon^{\alpha\beta}\right)\,,
\end{split}
\end{equation}
where the factors match the same conventions as in \eqref{eq:dMNP}. The wedge product \eqref{eq:wedge27-def} decomposes as
\begin{equation}
\begin{split}
(J_u\wedge J_v)_{\mbf{ab},\alpha\beta}&=\frac{1}{\sqrt{10}}\Bigl(\epsilon_{\mbf{abcde}} J_{(u}{}^{\mbf{cd}} J_{v)}{}^{\mbf{de}}{}_{\alpha\beta}-4\, J_{u,[\mbf{a}|,[\alpha|} J_{v,|\mbf{b}],|\beta]})\Bigr)\,,\\ 
(J_u\wedge J_v)_{\mbf{a}}&=\sqrt{\frac{2}{5}}\Bigl(\frac{1}{8}\epsilon_{\mbf{abcde}}J_{u}{}^{\mbf{bc}}J_{v}{}^{\mbf{de}}-J_{(u}{}^{\alpha}J_{v),\mbf{a},\alpha}\Bigr)\,,\\
(J_u\wedge J_v)^{\mbf{a}}{}_\alpha&=-\sqrt{\frac{2}{5}}\,\Bigl(J_{(u}{}^{\mbf{ab}} J_{v),\mbf{b},\alpha}+J_{(u}{}^\gamma J_{v)}{}^{\mbf{a}}{}_{\gamma\alpha}\Bigr)\,,\\
(J_u\wedge J_v)_{\alpha\,[\beta\gamma]}&=2\sqrt{\frac{2}{5}}\Bigl(J_{(u|}{}^{\mbf{a}}{}_{\alpha[\beta|} J_{|v),\mbf{a},|\gamma]}\Bigr)\,,
\end{split}
\end{equation}
and the generalised Lie derivatives \eqref{eq:GenLieDerivative} as
\begin{equation}
\begin{split}
(\mathcal{L}_{J_u}J_v)^{\mbf{ab}}&=\mathcal{L}_{(J_u)_{\textbf{10},(0)}}(J_v)^{\mbf{ab}}+L_{(J_u)_{\textbf{0},(v)}}(J_v)^{\mbf{ab}}-L_{(J_v)_{\textbf{0},(v)}}(J_u)^{\mbf{ab}}+J_v{}^\alpha (\mathfrak{d}(J_{u,\bar{\textbf{5}},(1)}))^{\mbf{ab}}{}_\alpha \\
&\qquad +\frac{1}{2}\epsilon^{\mbf{abcde}}\partial_{\mbf{cd}}J_u{}^\alpha J_{v,\mbf{e},\alpha}\,,\\
(\mathcal{L}_{J_u}J_v)^{\mbf{a}}{}_{\alpha\beta}&=\mathcal{L}_{(J_u)_{\textbf{10},(0)}}(J_v)^{\mbf{a}}{}_{\alpha\beta}+L_{(J_u)_{\textbf{0},(v)}}(J_v)^{\mbf{a}}{}_{\alpha\beta}+2\, J_v{}^{\mbf{ab}}\partial_{[\alpha}J_{u,\mbf{b},|\beta]}+2\,\partial_{[\alpha}J_u{}^{\mbf{ab}} J_{v,\mbf{b},|\beta]}\\
&\qquad-(J_v){}^{\mbf{ab}}(\mathfrak{d}(J_{u,\textbf{5},(2)}))_{\mbf{b},\alpha\beta}+2\,(\mathfrak{d}(J_{u,\bar{\textbf{5}},(1)}))^{\mbf{ab}}{}_{[\alpha|} J_{v,\mbf{b},|\beta]}\,,\\
(\mathcal{L}_{J_u}J_v)_{\mbf{a},\alpha}&=\mathcal{L}_{(J_u)_{\textbf{10},(0)}}(J_v)_{\mbf{a},\alpha}+L_{(J_u)_{\textbf{0},(v)}}(J_v)_{\mbf{a},\alpha}-\frac{1}{4}\epsilon_{\mbf{abcde}}J_v{}^{\mbf{bc}}(\mathfrak{d}\,(J_{u,\bar{\textbf{5}},(1)}))^{\mbf{de}}{}_\alpha-\partial_{\mbf{ab}} J_u{}^\beta J_v{}^{\mbf{b}}{}_{\beta\alpha}\\
&\qquad +J_v{}^\beta (\mathfrak{d}(J_{u,\textbf{5},(2)}))_{\mbf{a},\beta\alpha}-J_v{}^\beta\partial_\beta J_{\mbf{a},\alpha}+J_v{}^\beta\partial_\alpha J_{\mbf{a},\beta}-\frac{1}{4}\epsilon_{\mbf{abcde}}\partial_\alpha J_u{}^{\mbf{bc}} J_v{}^{\mbf{de}}\,,\\
(\mathcal{L}_{J_u}J_v)^\alpha&=\mathcal{L}_{(J_u)_{\textbf{10},(0)}}(J_v)^\alpha+L_{(J_u)_{\textbf{0},(v)}}(J_v)^\alpha-\mathcal{L}_{(J_v)_{\textbf{10},(0)}}(J_u)^\alpha\,,
\end{split}
\end{equation}
where $\mathcal{L}$ in the right hand side is the $\SL{5}$ generalised Lie derivatives and $L$ is the usual Lie derivative along the GL(2) directions. The operator $\mathfrak{d}$ is the $\SL{5}$ ExFT nilpotent differential operator \cite{Wang:2015hca,Malek:2017njj} and $d$ is the usual exterior derivative with respect to the GL(2) coordinates.

\providecommand{\href}[2]{#2}\begingroup\raggedright\endgroup

\end{document}